\documentclass[10pt,twocolumn,twoside]{IEEEtran}
\usepackage{graphicx}
\usepackage{amsmath}
\usepackage{amsfonts}
\usepackage{amssymb}
\usepackage{algorithm}
\usepackage{algorithmic}
\usepackage{epsfig}
\usepackage{url}
\newcommand{\beq}{\begin{equation}}
\newcommand{\eeq}{\end{equation}}
\newcommand{\bec}{\begin{center}}
\newcommand{\eec}{\end{center}}
\newcommand{\barr}{\begin{array}}
\newcommand{\earr}{\end{array}}
\newcommand{\beqnarr}{\begin{eqnarray}}
\newcommand{\eeqnarr}{\end{eqnarray}}
\newcommand{\beqarr}{\begin{eqnarray*}}
\newcommand{\eeqarr}{\end{eqnarray*}}

\newcommand{\hs}[1]{\mbox{\hspace{#1}}}
\def\picture #1 by #2 (#3){
  \vbox to #2{
    \hrule width #1 height 0pt depth 0pt
    \vfill
    \special{picture #3} % this is the low-level interface
    }
  }

\def\scaledpicture #1 by #2 (#3 scaled #4){{
  \dimen0=#1 \dimen1=#2
  \divide\dimen0 by 1000 \multiply\dimen0 by #4
  \divide\dimen1 by 1000 \multiply\dimen1 by #4
  \picture \dimen0 by \dimen1 (#3 scaled #4)}
  }

\title{Bayesian Segmentation of Oceanic SAR Images: Application to Oil Spill Detection}

\author{S\'{o}nia~Pelizzari and Jos\'{e} M. Bioucas-Dias
\thanks{This work  was
       supported by the Funda\c{c}\~{a}o para a Ci\^{e}ncia e Tecnologia, under
       the grant PDCTE/CPS/49967/2003 and by the European Space
       Agency, under the grant ESA/C1:2422/2003.}
\thanks{The authors are with Instituto de Telecomunica\c{c}\~{o}es and
        Instituto Superior T\'{e}cnico, Av. Rovisco Pais, Torre Norte,
        Piso 10, 1049-001 Lisboa, Portugal (email:\{bioucas, soniap\}@lx.it.pt,
        tel:+35121841846\{6,7\}, fax:+351218418472).}
}
\begin{document}
\maketitle

\begin{abstract}
This paper introduces Bayesian supervised and unsupervised segmentation
algorithms aimed at oceanic segmentation of SAR images.  The data term, \emph{i.e}.,
the density of the observed backscattered signal given the region, is modeled
by a finite mixture of Gamma densities with a given predefined number of
components. To estimate the parameters of the class conditional densities, a
new expectation maximization algorithm was developed. The prior is a
multi-level logistic Markov random field  enforcing local continuity in a
statistical sense. The smoothness parameter controlling the degree of
homogeneity imposed on the scene is automatically estimated, by computing the
evidence with loopy belief propagation; the classical coding and least squares
fit methods are also considered. The maximum a posteriori  segmentation is
computed efficiently by means of recent graph-cut techniques, namely the
$\alpha$-Expansion algorithm that extends the methodology to an optional number
of classes. The effectiveness of the proposed approaches is illustrated with
simulated images and real ERS and Envisat scenes containing oil spills.
\end{abstract}


% File Name         : OILSAR_article_sec1.tex
% Contents          : sec1
% Paper Title       : Oil Spill Segmentation
%                   :
%
% Paper ID          :
% Person ID         :
% Date              : 01/11/07
% Author            : Jose M. Bioucas-Dias and S\'{o}nia Pelizzari

\section{Introduction}\label{sec:1}
A wide number of oceanic  phenomena become visible 
on SAR images as they have distinct scattering characteristics, namely the sea surface roughness and thus the normalized radar cross section (NRCS). Among these phenomena are gravity waves, convective cells, oceanic internal waves, current and coastal fronts, eddies, upwelling processes, ship wakes and oil pollution \cite{book:NOAA:seamanual}. The automatic detection of these signatures is of outmost interest for a panoply of ocean monitoring systems, both for security, for commercial and for research applications. An example of an automated ocean feature detection scheme, able to detect fronts, ice edges and polar lows, is  described in \cite{art:Wu:IJRS:03}. Another example is the Ocean Monitoring Workstation (OMW), developed by the company Satlantic.
Usually, the first two steps  of the   processing chains of such systems 
are segmentation followed by  classification. The Segmentation  step  computes a set of regions defining an
image partition, where the features of each region, for example the gray levels,  are {\em similar} in some sense. 
The classification focus then on each region  attaching a label to it. For example, oil spill detection based on SAR  images
has bee approached by many authors with  the referred to classification-segmentation scheme(see for example \cite{art:Brekke:OSD:07}); the SAR image
is first segmented and then the classification is focused on the regions with lower scattering; the classifier then computes 
a number of features from these regions including shape, moments, scale parameters, etc. \cite{art:Brekke:OSD:07} based on which a
decision on  whether the region corresponds to oil or look-alike is taken. We should refer, however,

 An  application where segmentation of oceanic SAR images is important is oil spill detection. In fact, many approaches to this issue have been proposed in
recent years. Most of them follow the described segmentation-classification structure, although other methods exist that do not, like for example kernel-based anomaly detectors \cite{art:Mercier:TGRS:06}. A review of SAR segmentation techniques for oil spill detection can be found in
\cite{art:Brekke:OSD:07}. The present work, which is an elaboration of our
previous works \cite{conf:Pelizzari:SeaSAR:06},
\cite{conf:Pelizzari:IbPRIA:07}, \cite{conf:Pelizzari:IGARSS:07}, describes
algorithms for the segmentation of dark signatures in oceanic SAR images, following a Bayesian
approach. The adopted  technique uses as data model a finite Gamma mixture,
with a given predefined number of components, for modeling each class density.
In fact, this density is well suited to filtered intensity SAR images as shown,
\emph{e.g.}, in \cite{conf:Bioucas:ICIP:99} and in \cite{art:Delignon:1997}. By using a mixture, we aim at
describing the continuous backscattering variability that may be observed in
the SAR sea data. Moreover, a mixture is able to describe densities presenting more than one maximum, as it is the case of oceanic multi-look SAR images histograms. This cannot be achieved by none of the statistical models for SAR intensities proposed in the literature,  like for example any of the Pearson distributions.

In this work we propose two supervised and one unsupervised algorithm. The supervised algorithms demand an
interaction with the user that manually selects a region containing pixels from the dark signature of interest
and a region containing water pixels. These regions should be representative and can be made up of different not connected parts. The unsupervised
algorithm is an improvement of the supervised ones and is completely automatic.

To estimate the parameters of the class conditional
densities, a new expectation maximization (EM) algorithm was developed. Details
are described in appendix. When segmenting small sub-scenes of the image, a
simplified data model with only one Gamma function per class can be used.

The prior used to impose local homogeneity is a multi-level logistic (MLL)
 model Markov random
field (MRF) \cite{book:Li:ComputerVision}, with 2nd order neighborhood. To
infer the prior smoothness parameter controling the degree of scene
homogeneity, we develop an EM algorithm that uses loopy believe propagation
(LBP) \cite{conf:Yedidia:IJCAI:01}.  We have also exploited different classic
estimation methods, namely the least squares fit (LSF) and the coding method
(CD) (see \cite{book:Li:ComputerVision} for details) for comparison.

To infer the labels, we adopt the maximum a posterior (MAP) criterion, which we
implement efficiently and exactly with graph-cut techniques
\cite{art:Kolmogorov:PAMI:04}.

Although the segmentation of oceanic dark patches is
typically based on the assumption of two classes, we have
generalized the problem to an optional number of classes. The underlying
integer optimization problem is now  attacked with the graph-cut based
$\alpha$-Expansion algorithm \cite{art:Boykov:V:Z:PAMI:01}.

To evaluate the accuracy of
the proposed algorithms, different simulations addressing both the referred
Gamma model as well as intensity images corrupted with Gaussian noise have been
carried out for error rate assessment.

The algorithms have also been applied to
real SAR images containing well documented oil spills. For doing so, the scenes are divided in tiles and segmented individually.

\subsection{Related Work}
\label{sec:1.1}

Related approaches to the problem of oil spill segmentation are built on
off-the-shelf segmentation algorithms such as the \emph{adaptive image
thresholding} and the  \emph{hysteresis thresholding}. Entropy  methods based
on the maximum descriptive length (MDL) and wavelet based approaches have also
been proposed. Another recently proposed segmentation methodology applies Hidden Markov Chains (HMC) to a multiscale representation of the original image. Hereby the wavelet coefficients are statistical characterized by the Pearson system and by the the generalized
Gaussian family \cite{art:Derrode:PR:07}.
When other SAR products are available, for example polarimetric data, other methods have been described in the literature, like constant false alarm
rate filters \cite{art:Maurizio:TGRS:07}.

An example of an elaborated adaptive thresholding technique is provided by
\cite{art:Solberg:OSDIR:07}. In this method, an image pyramid is created by
averaging pixels in the original image. From the original image, the next level
in the pyramid is created with half the pixel size of the original image. A
threshold is then computed for each level based on local estimates of the
roughness of the surrounding sea and on a look-up table containing experimental
values obtained from a training data set.

Hysteresis thresholding has been used as the base for detecting oil slicks in
\cite{conf:Kanaa:IGARSS:03}. The method includes two steps: applying a
so-called directional hysteresis thresholding (DHT) and performing the fusion
of the DHT responses using a Bayesian operator. The MDL technique, which  basically
consists in applying information theory in order to find the image description
which has the lowest complexity,  has been applied in
\cite{art:Galland:SAR_OSS:04}  to segment speckled SAR images,  namely
those containing oil slicks. This segmentation method describes the image as a
polygonal grid and determines the number of regions and the location of the
nodes that delimit the regions. The two-dimensional wavelet transform, used as
a bandpass filter to separate processes at different scales, has also been
adopted to oil slick detection in the framework of an algorithm for automated
detection and tracking of mesoscale features from satellite imagery.
\cite{art:Wu:IJRS:03}.

\subsection{Contributions}
\label{sec:1.2}

We approach oil spill segmentation using a Bayesian framework and a
multi-level Logistic (MLL) prior. Several methods in the same vein have been
proposed since the seminal work of Geman and Geman \cite{art:Geman:IEEETPA:84},
see \emph{e.g.}, \cite{art:Berthod:IVC:96}. Applications of these ideas in the
segmentation of SAR images can be found, \emph{e.g.}, in
\cite{art:Kelly:TASSP:88}, \cite{art:Derin:TGRS:90},
\cite{art:Dellepiane:TGRS:97}.

The main contributions of this work to the state-of-art in oil spill
segmentation are the following:
\begin{itemize}
\item the development of an EM algorithm to estimate the parameters of a  mixture
of a pre-defined number of Gamma distributions, in order to model the intensities
in a SAR image

\item the development of an EM algorithm using LBP to estimate the smoothness
parameter in the MRF used as pior in our framework

\item the application of recent graph-cut techniques for solving the energy
minimization problem that arises from the followed Bayesian methodology.

\item the design of a semisupervised algorithm for oil spill segmentation
supported on the tools referred to above.

\end{itemize}

\subsection{Paper Organization}
The article is organized as follows: Section 1 introduces the problem, with
references to related work and giving the main contributions of the present
work; Section 2 overviews the Bayesian methodology that builds the base to the
proposed algorithms. In addition, it briefly reviews the concept of the
$\alpha$-Expansion technique that has been implemented to generalize the
methodology to an optional number of classes; Section 3 describes the used
parameter estimation techniques; Section 4 describes the supervised segmentation
algorithms; Section 5 describes the unsupervided algorithm and Section 6 presents results of segmenting simulated and real images applying the algorithms proposed in Section 4 and 5;
Finally Section 7 contains the main conclusions and future work remarks. The
article also includes an appendix where the referred  to EM algorithm,
developed to estimate the parameters of the class conditional densities of the
Gamma mixture data model, is described.
\label{sec:1.3}

% File Name         : OILSAR_article_sec2.tex
% Contents          : sec2
% Paper Title       : Oil Spill Segmentation
%                   :
%
% Paper ID          :
% Person ID         :
% Date              : 01/11/07
% Author            : Jose M. Bioucas-Dias and S\'{o}nia Pelizzari

\section{Problem Formulation}\label{sec:2}
\subsection{Bayesian Approach}\label{sec:2.1}

let ${\cal L}:=\left\{1,\ldots, c\right\}$ be a set of $c$  classes and  ${\cal
P}:=\left\{1, 2, \ldots, N\right\}$ be the set of $N$  pixels (sites) where
measurements $y :=\left\{y_{1}, y_{2}, \ldots, y_{N}\right\}$, the SAR
intensities, are available. A labeling $x:=\left\{x_{1},x_{2},\ldots
x_{N}\right\}$ is a mapping from ${\cal P}$ to ${\cal L}$, {\em i.e.}, it
assigns to each pixel $p \in {\cal P}$ a label $x_{p} \in {\cal L}$. Any
labeling $x$ can be uniquely represented by a partition of image pixels $P =
\left\{{P_{l} | l \in {\cal L}}\right\}$, where $P_{l} = \left\{ p \in  {\cal
P}| x_{p} = l\right\}$ is the subset  of pixels to which the label $l$ have
been assigned. Since there is an one-to-one correspondence between labelings
$x$ and partitions $P$, we  use these notions interchangeably. By applying a
segmentation algorithm to the image y, we get
$\hat{x}:=\left\{\hat{x}_{1},\hat{x}_{2} \ldots \hat{x}_{N}\right\}$, where
$\hat{x}_{i}$ is the  inferred label for pixel $i \in {\cal P}$.

\subsection{Observation Model}

In our problem formulation, we assume conditional independence of the
measurements  given the labels, {\em i.e.},
\begin{equation}
p(y|x)=\prod_{i=1}^{N}p(y_{i}|x_{i}) = \prod_{l=1}^c\prod_{i\in {\cal P}_l}
p(y_i|\phi^l),
\label{eq:BAYESIAN1}
\end{equation}
where $p(\cdot|\phi^l)$  is the density  corresponding to class $l$ and
$\phi_l$ the correspondent vector of parameters.  The adopted density is a
finite Gamma mixture given by
\begin{equation}
p(y_{i}|\phi^l)=\sum_{s=1}^{K}\alpha_{s}^l\, p(y_{i}|\theta_s^l),
\label{eq:BAYESIAN2}
\end{equation}
where $K$ is the number of Gamma modes in the mixture, $i$ indexes the pixel,
and, for the class $l$, $\theta_s^l$ is the vector of parameters of the Gamma
mode $s$, and $\alpha_s^l$ is the  a priori probability of mode $s$. 
 We denote
$\theta^l := (\theta_1^l,\dots,\theta_K^l)$,
$\alpha^l:=\left(\alpha_{1}^l,\ldots,\alpha_{K}^l\right)$, and $\phi^l :=
(\alpha^l,\theta^l)$.

Given that $p(y_i|\theta_s^l)$ is Gamma distributed, we have then
\begin{equation}
p\left(y_{i}|\theta_{s}^l\right)=\frac{(\lambda_{s}^l)^{a_s^l}}
             {\Gamma\left(a_{s}^l\right)}y_{i}^{a_{s}^l-1}\exp{\left(-\lambda_{s}^ly_{i}\right)},
             \hspace{0.5cm y_i \geq 0},
\label{eq:BAYESIAN3}
\end{equation}
where  $\theta_{s}^l:=\left(a_{s}^l,\lambda_{s}^l\right)$. The mean and variance 
of a random variable with the density (\ref{eq:BAYESIAN3}) is, respectively, 
$a_{s}^l/\lambda_{s}^l$ and $a_{s}^l/(\lambda_{s}^l)^{2}$.

The mixture parameters $\theta^l$  are estimated from the
data by applying the EM Gamma mixture estimation algorithm described in Appendix.
The procedure is the same both for the supervised and for the unsupervised algorithms.

We  now  make a brief comment on our choice of the Gamma mixture  for modeling
the observation densities.  Under the assumption of fully developed speckle,  
the complex  radar amplitude is  zero-mean circular Gaussian distributed \cite{book:Jakowatz96}. 
The average intensity
computed over a number of  independent random variables with the same density 
is, thus, Gamma distributed.
It happens, however, that in sea SAR imaging one or more of the above assumptions may fail,
rendering the Gamma density a poor model  for SAR intensity \cite{art:Derrode:PR:07}. A  line
of attack  to obtain better models is to  use more flexible parametric families,  such as
the Pearson System \cite[Ch. 4.1]{art:Jonhnson:Kotz:94} (see also \cite{art:Delignon:1997}).
However, in our problem each class density  exhibits much more variability than that of 
accommodated by the   the Pearson System. We have very often, for example, multi-modal densities.
We should resort, therefore, to a mixture of  Pearson System densities, what would lead 
to  complex learning procedures.  We have experimentally observed, however, that  
the Gamma mixture yields very good fittings for  real SAR histograms,  obtained with a
moderately complex learning algorithm.  For this reason, we have  adopted the Gamma 
mixture model.

\subsection{Prior}

\label{sec:prior}

A second assumption we are making is of local smoothness of the labels in a 
statistical sense. It is more likely to have neighboring sites with the same label than
the other way around. We model this  local smoothness with a
second order MRF, $P\left(x\right)$;  more specifically, by  an MLL model
(Ising model in the case of two classes). The Markov property assumes that
\begin{equation}
p\left(x_{i}|~x_{j},j \in {\cal P}\right)=p\left(x_{i}|~x_{j},j \in {\cal
N}_{i}\right), \label{eq:BAYESIAN4}
\end{equation}
where ${\cal N}_{i}$ is the set of neighbors of site $i$.  If
$p\left(x_{i}|~x_{j},j \in {\cal N}_{i}\right)>0$, then the Hammersley-Clifford
Theorem states that $p(x)$ has the form of the Gibbs distribution
\begin{equation}
p\left(x\right)=\frac{1}{{\cal Z}}\exp^{-U\left(x\right)}, \label{eq:BAYESIAN5}
\end{equation}
where ${\cal Z}$ is the so called partition function and $U$ is the energy function
\begin{equation}
U\left(x\right)=\sum_{cl \in C}V_{cl}\left(x\right), \label{eq:BAYESIAN6}
\end{equation}
where $C$ is the set of cliques and $V_{cl}\left(x\right)$ is the clique
potential defined over clique $cl$. In this work, we have pair-wise cliques
defined on a second-order neighborhood (8 pixels). That is, $C = \{ (i,j)\, :
 i\in{\cal N}_j,\; j\in{\cal N}_i, \; i>j\}$.

In these conditions, the  MLL  clique potentials, in the isotropic case,
is given by
\begin{equation}
V_{cl}\left(x_{r},x_{s}\right)=-\beta\delta\left(x_{r}-x_{s}\right),
\label{eq:BAYESIAN7}
\end{equation}
where $r,s\in cl$,  $\delta(x) := I_{0}(x)$  is the indicator function of set
$\{0\}$, and parameter $\beta>0$ controls the degree of scene homogeneity.

\subsection{Maximum a Posteriori Estimate}

The posterior of the labeling given the observed data is
\begin{equation}
p\left(x|y\right)\propto p\left(y|x\right)p\left(x\right). \label{eq:BAYESIAN8}
\end{equation}
In order to infer $x$, we adopt the MAP criterion. This amounts to maximize the
posterior density of the labeling given the observed data:
\begin{equation}
\hat{x}= \arg\max_{x} p\left(x|y\right), \label{eq:BAYESIAN9}
\end{equation}
and is equivalent to minimize the negative logarithm of (\ref{eq:BAYESIAN8}).
In this sense, we may rewrite the problem in the following way:
\begin{equation}
\hat{x}= \arg\min_{x} E(x_1,\dots,x_n),
\end{equation}
where
\begin{equation}
E\left(x_{1},\ldots,x_{N}\right) = -\log p(x|y) + c^{te}\label{eq:ENERGY1},
\end{equation}
and $c^{te}$ denotes  an irrelevant constant. From equation
(\ref{eq:BAYESIAN8}), we have
\begin{equation}
E\left(x_{1},\ldots,x_{N}\right)=\sum_{p=1}^{N}E^{i}\left(x_{i}\right)+\sum_{i,j\in
cl}E^{i,j}\left(x_{i},x_{j}\right), \label{eq:ENERGY2}
\end{equation}
with
\begin{eqnarray}
E^{i}\left(x_{i}\right) & = & -\log p\left(y_{i}|x_{i}\right)\\
              E^{i,j}(x_i,x_j) & = & -\beta\delta(x_i-x_j).
\label{eq:LIKELIHOOD1}
\end{eqnarray}

\subsection{Energy Minimization}\label{sec:2.2}

As already stated, we are concerned with the minimization of
$E\left(x_{1},\ldots x_{N}\right)$ given by (\ref{eq:ENERGY2}), which we term
energy. For two classes, (\emph{i.e.}, $c=2$),  the global minimum of
$E\left(x_{1},\ldots x_{N}\right)$ can be computed exactly by applying the
graph-cut algorithm described in \cite{art:Kolmogorov:PAMI:04}. This is a
consequence of energy being graph-representable, \emph{i.e.},
$E^{i,j}\left(0,0\right)+E^{i,j}\left(1,1\right)\leq
E^{i,j}\left(0,1\right)+E^{i,j}\left(1,0\right)$ (for details see
\cite{art:Kolmogorov:PAMI:04}).  For more than two classes, the solution of
(\ref{eq:BAYESIAN9}) can be approximately computed by the
 $\alpha$-Expansion technique \cite{art:Boykov:V:Z:PAMI:01}, also based on
graph-cut concepts. This algorithm finds the local minimum of the energy within
a known factor of the global minimum.

We now give a brief  description of the $\alpha$-Expansion algorithm. Given a
label $\alpha$, a move from a partition $P$ (with correspondent labeling $x$)
to a new partition $P'$ (with correspondent labeling $x'$) is called an
$\alpha$-Expansion if $P_{\alpha} \subset P'_{\alpha}$ and $P'_{l} \subset
P_{l}$ for any label $l \neq \alpha$. In other words, an $\alpha$-Expansion
move allows any set of image pixels to change their label to $\alpha$. The
algorithm cycles through the labels in ${\cal P}$ in some fixed or random order
and finds the lowest $\alpha$-Expansion move from the current labeling. If the
expansion move has lower energy than the current labeling, then it becomes the
current labeling. The algorithm terminates when no $\alpha$-Expansion move
exist corresponding to a local minimum of the  energy.

% File Name         : OILSAR_article_sec3.tex
% Contents          : sec3
% Paper Title       : Oil Spill Segmentation
%                   :
%
% Paper ID          :
% Person ID         :
% Date              : 01/11/07
% Author            : Jose M. Bioucas-Dias and S\'{o}nia Pelizzari

\section{Estimation of  Parameter $\beta$}\label{sec:3}

In this work, we consider three different  techniques to determine the
smoothness parameter $\beta$: a new EM method hereafter introduced that uses
loopy belief propagation (LBP) and the classical LSF and CD methods. Because
LSF and CD assume the existence of labeled data,  we have conceived an
iterative labeling-estimation scheme, which alternates between a labeling step
and an estimation step until the convergence of $\beta$ is attained. On the
contrary, the EM estimation algorithm, that we have called
``Loopy-$\beta$-Estimation", is a one-shot technique. In the following
sections, we briefly review the LSF and CD methods and provide a detailed
description of the  Loopy-$\beta$-Estimation method. In this section,
the class parameters $\phi
:= (\phi^1,\phi^2,\dots,\phi^c)$ are assumed known.

\subsection{Least Squares Fit}\label{sec:3.1}

This procedure for parameter estimation in MRF, described in detail in
\cite{book:Li:ComputerVision}, is based in the following equation that holds
for the MLL model, for every pixel $p$, with neighborhood ${\cal N}_{p}$, and
for every label pair $x_p,x_p' \in {\cal L}:$
\begin{eqnarray}
\beta\left[n\left(x_p\right)-n\left(x_p'\right)\right] & = &
\log{\left(\frac{p\left(x_p|x_{{\cal N}_{p}},y\right)}{p\left(x_p'|x_{{\cal
N}_{p}},y\right)}\right)}\nonumber\\
         &&-\left(E^{p}\left(x_p'\right)-E^{p}\left(x_p\right)\right),\hspace{1cm}
\label{eq:LSF1}
\end{eqnarray}
where $n(x_p)$ is the number of pixels in the set ${\cal N}_p$ with the same
label as $x_p$ and $x_{{\cal N}_p}:=\{x_i,\,i\in {\cal N}_p\}$. We use
histograms to estimate  the joint probabilities  $p\left(x_p|x_{{\cal
N}_{p}},y\right)$ and  $p\left(x_p'|x_{{\cal N}_{p}},y\right)$  in
(\ref{eq:LSF1}): assuming that there are a total of $M$ distinct 3x3
blocks in the image lattice with a given  label configuration $x_{{\cal N}_p}$,
then we take
\begin{equation}
\widehat p\left(x_{p}|x_{{\cal N}_{p}},y\right)=\frac{{\cal
H}\left(x_{p}|x_{{\cal N}_{p}},y\right)}{M},
\label{eq:LSF2}\\
\end{equation}
where ${\cal H}\left(x_{p}|x_{{\cal N}_{p}},y\right)$  is the number of times
that a particular 3x3 configuration $\left(x_{p}|x_{{\cal N}_{p}},y\right)$
occurs. The expression is then evaluated for a number of distinct combinations
of $x_p, x_p'$ and $x_{{\cal N}_{p}}$ in order to obtain an over-determined
linear system of equations that is solved in order to $\beta$.

\subsection{Coding Method}\label{sec:3.2}

In this method the key idea  (see \cite{book:Li:ComputerVision}) is to
partition the set ${\cal P}$ into sets ${\cal P}^{\left(k\right)}$, called
codings, such that no two sites in one ${\cal P}^{\left(k\right)}$ are
neighbors. In the present work, the  neighborhood ${\cal N}_{p}$ is of 2nd
order thus yielding four codings. As the pixels in ${\cal P}^{\left(k\right)}$
are not neighbors, the variables associated with these pixels, given the labels
at all other pixels, are mutually independent under the Markovian assumption.
The following simple product is thus obtained for the likelihood:
\begin{equation}
p^{(k)}\left(x|\beta,y\right)= \prod_{p \in {\cal
P}^{\left(k\right)}}p\left(x_{p}|x_{{\cal N}_{p}},\beta,y\right),
\label{eq:CD1}\\
\end{equation}

Maximizing (\ref{eq:CD1}) in order to $\beta$ gives the coding estimate  $\hat{\beta}^{(k)}$.
Although it is not clear how to combine the results optimally, the arithmetic
average, as suggested in \cite{book:Li:ComputerVision}, is an intuitive scheme
that was adopted in this work.

\subsection{Loopy-$\beta$-Estimation}\label{sec:3.3}

We are seeking $\hat{\beta}_{ML}=\arg\max_{\beta} p(y|\beta)$, the ML estimate
of the smoothness parameter $\beta$. Based on the fact that the marginal
density $p\left(y|\beta\right)$, the so-called evidence, is a sum over the
missing labels $x$, \emph{i.e.},
\begin{eqnarray}
\lefteqn{p\left(y|\beta\right) \equiv \sum_{x}p\left(y,x|\beta\right)}\nonumber\\
&& \hs{0,65cm}
= \sum_{x}p\left(y|x\right)p\left(x|\beta\right)\label{eq:E(MISSL)},\\
\nonumber
\end{eqnarray}
we develop an EM algorithm \cite{art:Dempster:MLE_EM:77}  for the ML estimation
of the  parameter $\beta$. The EM algorithm alternates between two steps: the
E-step computes the conditional expectation of the logarithm of the complete a
posteriori probability function, with respect to the missing variables, based
on the actual parameter value; the M-step updates the value of the parameter,
by maximizing the expression obtained in the E-step with respect to that
parameter. We now derive the E-step and the M-step.
\newline

\noindent \textbf{E-step:}
\begin{eqnarray}
    Q(\beta; \beta_{t}) & = & E[\log p(y,x|\beta)|y,\beta_{t}]\\
                   & = & E[\log p(y|x)|y,\beta_{t}]
                   \label{eq:classTerm}\\
                    &&+ E[\log p(x|\beta)|y,\beta_{t}].
                       \label{eq:QBETA1}
\end{eqnarray}
Recalling  that the MLL prior is given by
\begin{equation}
p\left(x|~\beta\right)= \frac{1}{{\cal
Z}\left(\beta\right)}\exp{\left[\beta\sum_{i,j\in
cl}\delta\left(x_{i}-x_{j}\right)\right]},
\label{eq:MLL_ISO}
\end{equation}
with
\begin{equation}\label{eq:ZEXPRESSION}
{\cal Z}\left(\beta\right)=\sum_{x}\exp{\left[\beta\sum_{i,j\in
cl}\delta\left(x_{i}-x_{j}\right) \right]},
\end{equation}
we obtain, up to an irrelevant constant,
\begin{equation}
Q\left(\beta;\beta_{t}\right)= -\log{{\cal Z}\left(\beta \right)}+\beta
\sum_{i,j\in cl} E\left[\delta\left(x_{i}-x_{j}\right) |~y,\beta_{t}\right]
\label{eq:QBETA2}.\\
\end{equation}

\noindent \textbf{M-step:}
\begin{equation}
\beta_{t+1}=\arg\max_{~\beta}Q\left(\beta,\beta_{t}\right).
\end{equation}
The stationary points of ${\cal Q}$ are the solution of
\begin{eqnarray}
\lefteqn{\qquad\frac{\partial Q}{\partial \beta}=
    -\frac{\partial \log{\left[{\cal Z\left(\beta\right)}\right]}}{\partial \beta}}
    \label{eq:MSTEP1}\\
&& \hs{0,3cm} +\sum_{i,j\in cl}\sum_{x_{i},x_{j}\in{\cal L}}\delta
\left(x_{i}-x_{j})p\left(x_{i},x_{j}|~y,\beta_{t}\right)\right)=0.
\nonumber
\end{eqnarray}

By introducing expression (\ref{eq:ZEXPRESSION})  into (\ref{eq:MSTEP1}) and if
we consider that $\delta\left(x_{i}-x_{j} \right)$ takes non-zero values only
for equal labels, we obtain
\begin{eqnarray}
\lefteqn{\frac{\partial Q}{\partial \beta}=
         \sum_{i,j\in cl}\sum_{k=1}^{c}p\left(x_{i}=k,x_{j}=k|~y,\beta_{t}\right)}\nonumber\\
&& \hs{0cm}
-p\left(x_{i}=k,x_{j}=k|~\beta_{t}\right)=0.
\label{eq:MSTEP2}
\end{eqnarray}

Since computing exact marginal distributions is  infeasible in our case, we
replace them by pseudo-marginals using BP. This approach has been successfully
applied in problems of approximate parameter learning in discriminative fields
\cite{conf:Kumar:ISCV:95}. BP is an efficient iterative algorithm in which
local messages are passed in graphical models.  For singly-connected (loop
free) pairwise MRFs, the two-node beliefs will correspond to the exact two-node
marginal probabilities. In our case, however, the graph that corresponds to the
MRF contains cycles, preventing  the basic BP algorithm to be applied. We
resort to a slightly  modified version called loopy belief propagation. In
practice, this algorithm has often delivered good results
\cite{conf:Yedidia:IJCAI:01}. We approximate the marginal probabilities with
the two-node beliefs $b_{i j}\left(x_{i},x_{j}\right)$ and $b_{i
j}^{y}\left(x_{i},x_{j}\right)$. These will provide approximations respectively
for the marginals $p\left(x_{i},x_{j}|\beta_t\right)$ and
$p(x_{i},x_{j}|y,\beta_t)$. By doing so, the M-Step of our EM algorithm is
given by the sum of the differences between the two node beliefs that take the
evidence $y$ into consideration and those that do not make use of $y$:
\begin{equation}
\frac{\partial Q}{\partial \beta}=  \sum_{i,j\in cl}\sum_{k=1}^{c}b_{i
j}^{y}\left(k,k\right)-b_{i j}\left(k,k\right)=0.
\label{eq:MSTEP3}\\
\end{equation}

\begin{figure}
  \centering
  \includegraphics[width=60mm,angle=0]{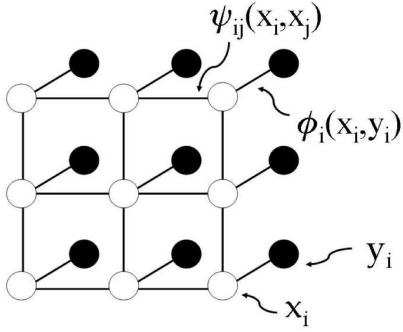}
  \caption{Lattice representing the pairwise MRF.}
  \label{fig:squareLatticeMRF}
\end{figure}

Fig. \ref{fig:squareLatticeMRF} depicts the graph that represents our pairwise
MRF for computing the two-node beliefs. In the square lattice,
$\psi_{ij}\left(x_{i},x_{j}\right)$ stands for the interaction potential that
penalizes every dissimilar pair of neighboring labels, and
$\phi\left(x_{i},y_{i}\right) = p\left(y_{i}|~x_{i}\right)$ represents the
statistical dependency between the labels $x_{i}$ and the measurements $y_{i}$.
For computing $b_{i j}\left(x_{i},x_{j}\right)$,
$\phi\left(x_{i},y_{i}\right)$, is set to a constant value, making the result
independent of the data values $y$.

We solve (\ref{eq:MSTEP3}) by a line search type algorithm,  ensuring that
$\frac{\partial^2 Q}{\partial^2 \beta}\le 0$, thus corresponding to a maximum
of $Q$.

\subsection{A few Remarks about the Vector of Parametres $\phi$}

In the previous section, we have considered that the class parameter
vector $\phi$ is known and only the smoothness parameter $\beta$ is to be
inferred. However, still using the beliefs computed by LBP, the vector $\phi$
could have been inferred simultaneously with $\beta$ by including
in the $Q$ function  the additive  term (\ref{eq:classTerm}) corresponding to
the class densities. We would have obtained
\begin{eqnarray*}
  Q(\phi;\phi_t,\beta_t) &=& E[\log p(y|x)|y,\phi_t,\beta_{t}] \\
             &=& \sum_{i=1}^N \sum_{l=1}^c \log p(y_i|\phi^l)p(x_i=l|y,\phi_t,\beta_{t}),
\end{eqnarray*}
where the probabilities $p(x_i=l|y,\phi_t,\beta_{t})$ are given by the LPB method.
The {\bf M-step} would consist then in two decoupled maximizations; one with respect to $\beta$
and another with respect to $\phi$. This approach is, however, beyond the scope
of this paper. Nevertheless, we present in Section \ref{sec:5} an
unsupervised algorithm, which is suboptimal but faster than the LBP based EM
algorithm aimed at
the inference of both the parameters $\beta$ and $\phi$.

An alternative to the proposed EM scheme based on LPB is using Monte Carlo techniques to cope with
the difficulty in computing the {\bf E-step} and the  partition function $Z$ \cite{art:Ibagnez:S:03},
\cite{art:Younes:89}, \cite{conf:Younes:92}. Supported on the performance of the LBP,
we believe, however, that the EM scheme based on LPB is, for the present problem,
much more faster  than the Monte Carlo based  techniques.

% File Name         : OILSAR_article_sec4.tex
% Contents          : sec4
% Paper Title       : Oil Spill Segmentation
%                   :
%
% Paper ID          :
% Person ID         :
% Date              : 01/11/07
% Author            : Jose M. Bioucas-Dias and S\'{o}nia Pelizzari

\section{Supervised Segmentation}\label{sec:4}

In this section, we introduce two supervised algorithms aimed at the
segmentation of sea SAR images. In both algorithms, the first step is the
estimation of the class parameters used in the data model. This is done by
asking the user to define representative regions of interest (ROI) in the
image. Once the regions are defined, different approaches may be followed: if a
Gamma mixture is assumed for the observed data, as in the case of SAR images,
the EM class parameters estimation algorithm described in Appendix is applied
to infer the class conditional densities. If we are segmenting small sub-scenes
and the radar range spreading loss has been compensated in the underlying SAR
images, one single Gamma function per class often provides a good modeling for
the SAR intensities. In this case, a common ML Gamma estimator is used instead
of the EM procedure. After this step, the data model is considered to be known
and is used thereafter.

The pseudo-code for two supervised algorithms is presented in Algorithm 1 and
Algorithm 2. Algorithm 1  is of generalized likelihood type
\cite{art:Lakshamanan:D:89} implementing an iterative labeling scheme with two
steps being performed alternately: the \textsl{$\beta$-Estimation} and the
\textsl{segmentation}. Algorithm 2 is a one-shot procedure that performs
$\beta$-Estimation using the Loopy $\beta$-Estimation method described in
Section \ref{sec:3.3}.

\begin{algorithm}
\label{alg1:segmentation} \caption{Supervised Segmentation Using LSF/CD~$\beta$-Estimation}
\begin{algorithmic}[1]
\REQUIRE Initial parameter $\hat{\beta}=\beta_{0}$ and estimated class
parameters $\hat{\phi}$
 \STATE Compute $eLabel=E^{p}(x_{p})$ for every pixel $p$ using $\hat{\phi}$
 \WHILE {$\left|\Delta\hat{\beta}\right|\leq \delta$ or nrIterations $<$ NrIterationsMax}
  \STATE Compute $\hat{x}=\alpha$-Expansion$(\hat{\beta},eLabel)$
  \STATE Compute $\hat{\beta}=\beta$-Estimation$(\hat{x},eLabel)$
 \COMMENT{LSF or CD}.
 \ENDWHILE
  \RETURN $\left(\hat{x},\hat{\beta}\right)$
 \end{algorithmic}
\end{algorithm}

\begin{algorithm}\label{alg2:segmentation}
\caption{Supervised Segmentation Using $\mbox{Loopy}$-$\beta$-Estimation}
\begin{algorithmic}[1]
\REQUIRE estimated class parameters $\hat{\phi}$
 \STATE Compute $eLabel=E^{p}(x_{p})$ for every pixel $p$ using $\hat{\phi}$
  \STATE Compute $\hat{\beta}= \mbox{Loopy}$-$\beta$-Estimation($eLabel$)
 \STATE Compute $\hat{x}= \mbox{GraphCut}$-Segmentation$(\hat{\beta},eLabel)$
  \RETURN $\left(\hat{x},\hat{\beta}\right)$
 \end{algorithmic}
\end{algorithm}

%Algorithm 2 is  currently working for 2 classes and uses the graph-cut
%technique described in \cite{art:Kolmogorov:PAMI:04}. Algorithm 1 is
%generalized for an optional number of classes, by adopting the
%$\alpha$-Expansion technique \cite{art:Boykov:V:Z:PAMI:01}.

% File Name         : OILSAR_article_sec5.tex
% Contents          : sec5
% Paper Title       : Oil Spill Segmentation
%                   :
%
% Paper ID          :
% Person ID         :
% Date              : 01/11/07
% Author            : Jose M. Bioucas-Dias and S\'{o}nia Pelizzari

\section{Unsupervised Segmentation}\label{sec:5}

The unsupervised method is an improvement of the supervised algorithms
described in the previous sections. The main difference is that the data
model is not considered to be known but is also iteratively estimated along
with the smoothness parameter and the segmentation. The scheme needs a rough
initialization of the data model parameters. We computed this initialization by fitting an EM Gamma mixture to
the complete data.

As our EM algorithm automatically eliminates unnecessary  modes, we  start with
an overestimate of  $K$, the number of modes in the mixtures. In our
experiments with real data, the maximum number of modes we got was four.

Then, different strategies are possible. In the experiments reported in the
next section, when considering $c=2$ classes, we have assigned the mode with a lower mean value to one of the classes and the remaining modes to the other class.

In each iteration, we compute the ML estimate of the vector $\phi$ based on the
previous segmentation, compute the new segmentation, and finally the new value
of $\beta$. The algorithm may use any of the three $\beta-$parameter estimation
methods and is applicable to an optional number of classes.
%Currently we can
%only use 2 classes when adopting Loopy-$\beta$-Estimation, due to
%implementation constrains, but the practical extention of the number of classes
%in this case in the future should be an easy task.
The algorithm stops when both the parameter $\beta$ and the class parameters
$\phi$ converge to stable values.

\begin{algorithm}
\label{alg3:segmentation} \caption{Unsupervised Segmentation}
\begin{algorithmic}[1]
\REQUIRE arbitrary parameter $\hat{\beta}=\beta_{0}$, initial class parameters
$\hat{\phi}=\phi_{0}$ (EM Gamma Mixture Estimation)
 \STATE Compute
$eLabel_{0}$\,=\,$E^{p}(x_{p})$ for every pixel $p$, using $\phi_{0}$ \STATE
Compute initial labeling \\
$\hat{x}=x_{0}=\alpha$-Expansion$(\beta_{0},eLabel_{0})$
 \FOR {stop criterium is not met }
  \STATE Compute $\hat{\phi}=$\,ML-Estimation$(\hat{x})$
  \STATE Compute $eLabel=E^{p}(x_{p})$ for every pixel $p$, using $\hat{\phi}$
  \STATE Compute $\hat{x}=\alpha$-Expansion$(\hat{\beta},eLabel)$
  \STATE Compute $\hat{\beta}=\beta$-Estimation$(\hat{x},eLabel)$
 \COMMENT{use LSF, CD or Loopy}.
 \ENDFOR
  \RETURN $\left(\hat{x},\hat{\beta},\hat\phi\right)$
 \end{algorithmic}
\end{algorithm}

% File Name         : OILSAR_article_sec6.tex
% Contents          : sec6 (two columns)
% Paper Title       : Oil Spill Segmentation
% Paper ID          :
% Person ID         :
% Date              : 8/7/05
% Author            : Jose M. Bioucas-Dias and S\'{o}nia Pelizzari

\section{Results}\label{sec:6}
This section presents results of applying the proposed methodology to simulated and real SAR images, as well as some considerations regarding time complexity of the algorithms.

\subsection{Simulations}\label{sec:6.1}
We have performed simulations corresponding to Gamma data terms
with two classes, and evaluated the overall accuracies of the obtained
segmentations using Algorithms 1,  2 and  3. We also illustrate in detail the EM Gamma mixture estimation algorithm, by applying it to an example of simulated data.

\subsubsection{Segmentation Results with Gamma Data Term}\label{sec:6.1.1}
\begin{figure}
  \centering
 \includegraphics[width=80mm,angle=0]{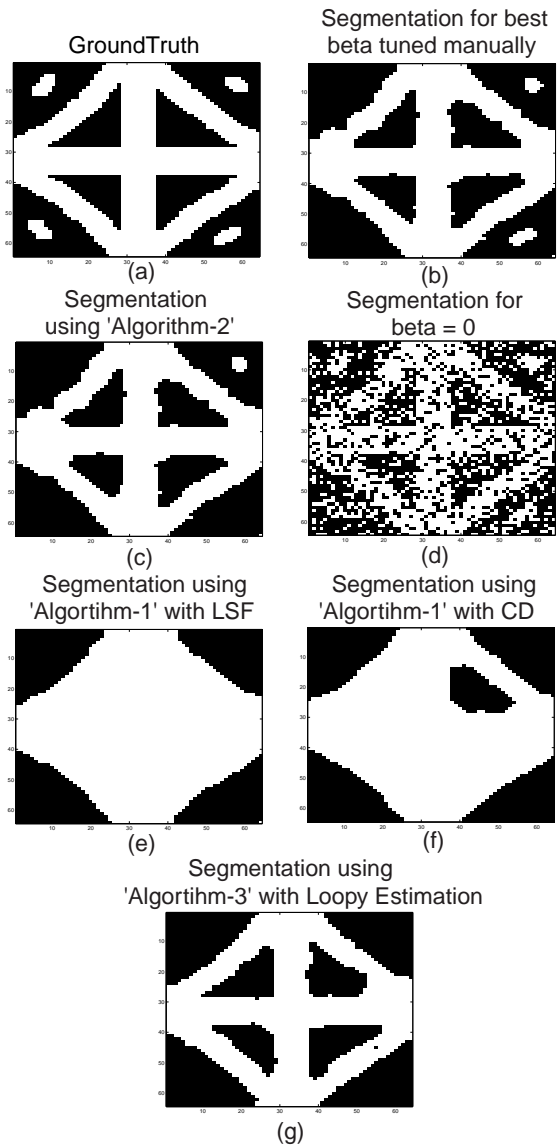}
  \caption{Ground-truth and results of different segmentation processes for an image with Gamma noise having mean values of five and nine and with a $\sigma$ value $=2.6$. Notice the good performance of Algorithm 2 and Algorithm 3, implementing the Loopy-$\beta$-Estimation. }
  \label{fig:AllSegmentationsSigma26}
\end{figure}

\begin{figure}
  \centering
  \includegraphics[width=90mm,angle=0]{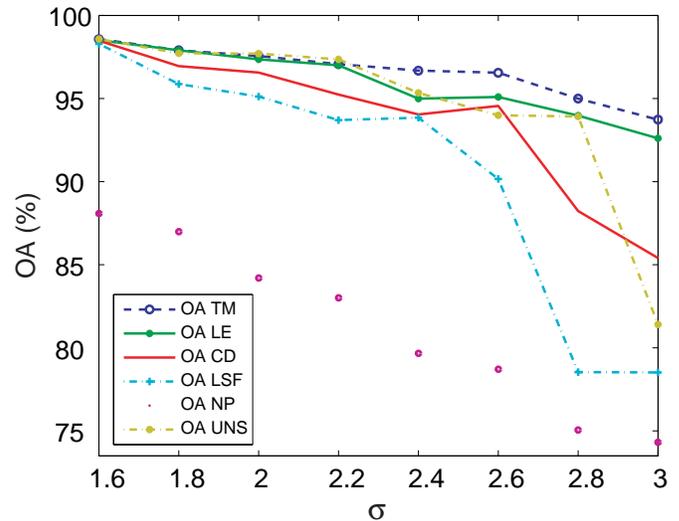}
  \caption{OA for images corrupted with Gamma noise with increasing $\sigma$ values: TM $\equiv$ for $\beta$ tuned manually; LE $\equiv$ Algorithm 2, using Loopy Estimation; LSF $\equiv$ Algorithm 1 using LSF; CD $\equiv$ Algorithm 1 using CD, NP $\equiv$ no prior and UNS $\equiv$ Algorithm 3 using Loopy Estimation.}
  \label{fig:OAVersusSigma}
\end{figure}

In order to compare the performance of the three proposed methods, we have tested Algorithms 1, 2 and 3 on simulated images generated by adding Gamma noise to ground-truth images containing two classes.
We have used the ground-truth depicted in Fig. \ref{fig:AllSegmentationsSigma26} (a) and added noise with Gamma distributions having mean values of five and nine. The parameters of the distributions were choosen in order to obtain increasing values of variance $\sigma^2$, corresponding to noisier images. We have then applied Algorithm 1,with LSF and CD methods, Algorithm 2, and Algorithm 3 with Loopy-$\beta$-Estimation. Furthermore, for comparison, the images were also segmented tuning the beta value manually and for the case that no prior is used ($\beta=0$). Fig. \ref{fig:AllSegmentationsSigma26} shows the segmentations obtained for $\sigma=2.6$. As we can see from $\beta=0$ this is a hard problem, on which LSF and CD fails the inner structures, but Loopy-$\beta$-Estimation provides better results, both in the supervised as in the unsupervised method. This behavior is further confirmed by the rank in Fig. \ref{fig:OAVersusSigma}, which gives the OA obtained for images with different $\sigma$ values (corresponding to images with more noise) for the six different segmentation processes referred above.

\begin{figure}
  \centering
  \includegraphics[width=90mm,angle=0]{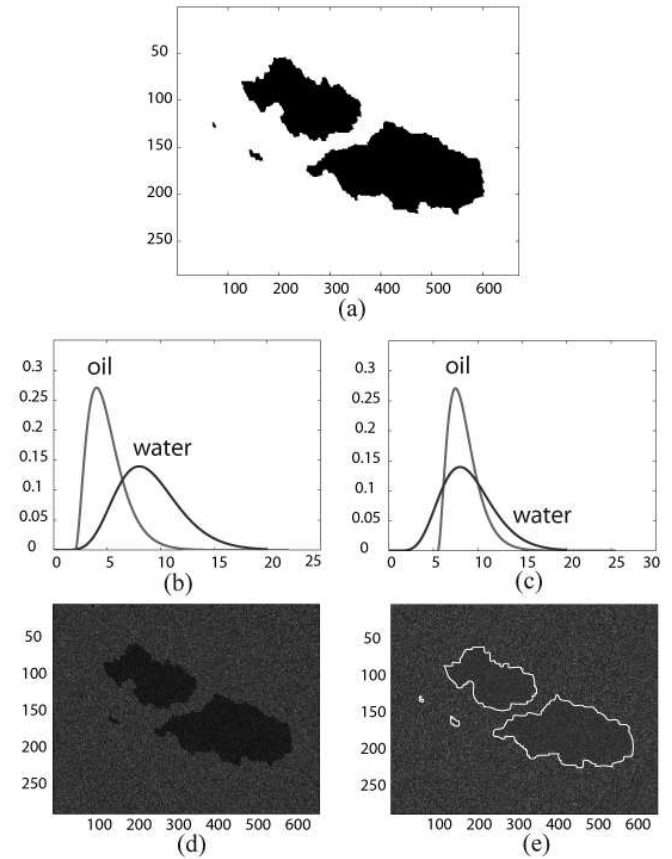}
  \caption{Gamma Model: (a) Ground-truth, (b) Histogram of Image A, (c) Histogram of Image B, (d) Image A, (e) Image B with superimposed delimiting line, for better visualization}
  \label{fig:GammaSimulation1}
\end{figure}

\begin{figure}
  \centering
  \includegraphics[width=110mm,angle=0]{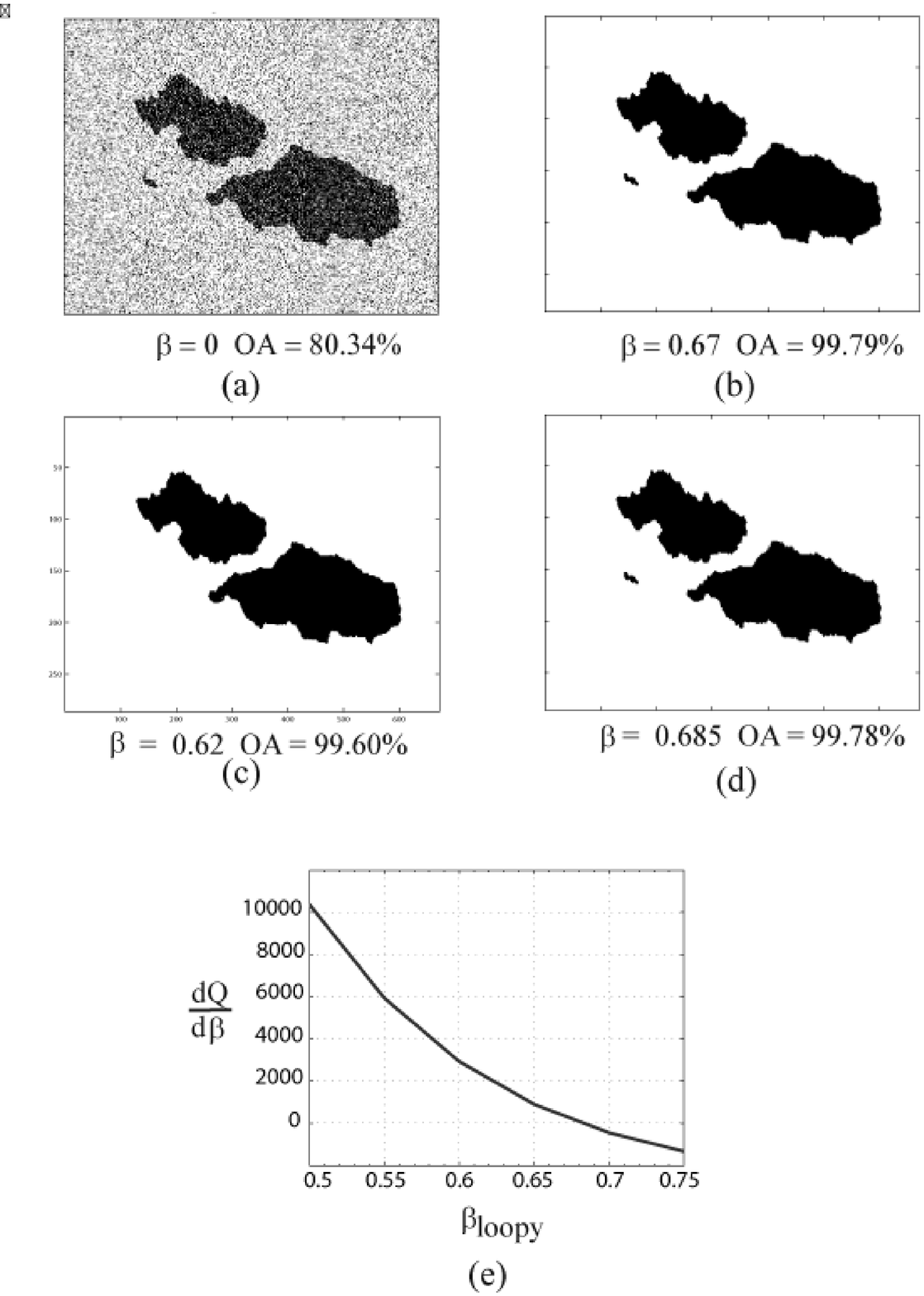}
  \caption{Segmentation results for image A: (a) No prior information, (b) $\beta$ value tuned manually, (c) Segmentation using Algorithm 3 with Loopy-$\beta$-Estimation, (d) Segmentation using Algorithm 2 (Algorithm 1 provided the same results with only 0.1\% difference in OA), (e) graph depicting the $\beta$ estimation in Algorithm 2.}
  \label{fig:GammaSimulation2}
\end{figure}

\begin{figure}
  \centering
  \includegraphics[width=85mm,angle=0]{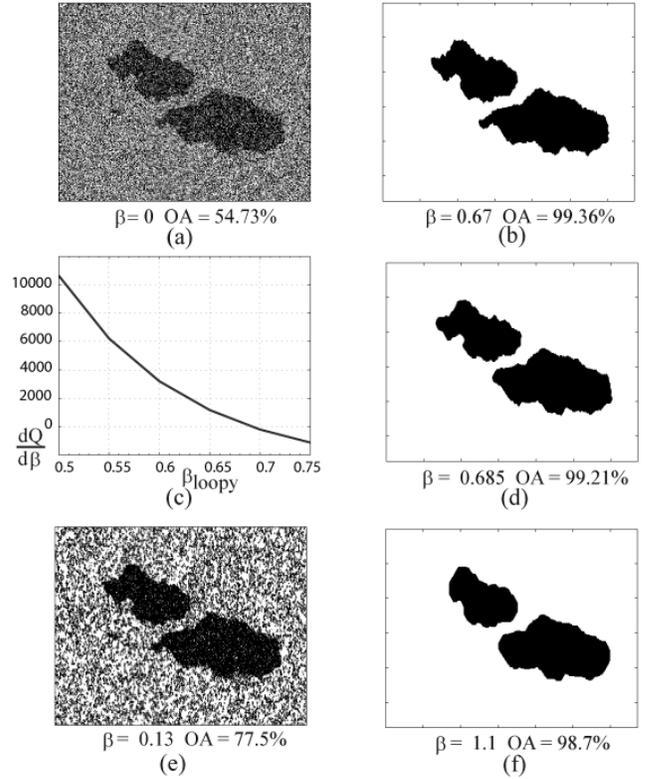}
  \caption{Segmentation results for Image B: (a) No prior information, (b) $\beta$ value tuned manually, (c) graph depicting $\beta$ estimation using Algorithm 2, (d) Segmentation using Algorithm 2, (e) Segmentation using Algorithm 1 with LSF and (f) Segmentation using Algorithm 1 with CD.}
  \label{fig:GammaSimulation3}
\end{figure}

We have also tested the proposed algorithms in images generated by adding Gamma noise to a ground-truth similar to an oil patch (see Fig. \ref{fig:GammaSimulation1}). Two images, Image A and Image B, with the histograms depicted in Fig. \ref{fig:GammaSimulation1}, corresponding to different segmentation difficulty levels are segmented. The best segmentation possible in this framework, achieved by tuning the $\beta$ value manually, and the segmentation obtained with no prior information (setting $\beta = 0$) are also displayed for comparison. Fig. \ref{fig:GammaSimulation2} shows the results obtained for Image A and Fig. \ref{fig:GammaSimulation3} the results obtained for Image B. For Image A, the segmented images using Algorithm 1, with CD and with LSF, are not shown, as they are almost equal to the image segmented with Algorithm 2. For Image B, Algorithm 3 has not provided good results and the segmentation is not displayed. The bad performance of Algorithm 3 in this case arises from the not so good initialisation of the class parameters, due to the complete overlapping from the oil and the water histogramms.

\hspace{1mm}

\subsubsection{EM Algorithm for Gamma Mixture}\label{sec:6.1.2}

\begin{figure}
  \centering
  \includegraphics[width=90mm,angle=0]{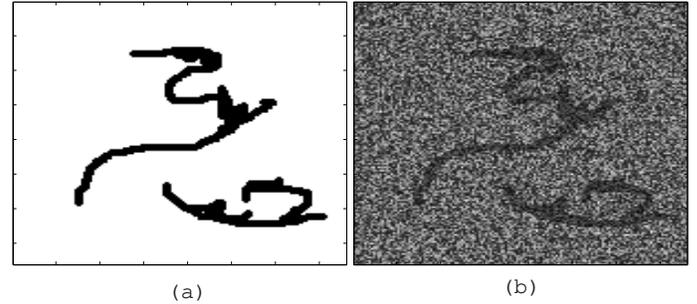}
  \caption{(a) Ground-truth used for simulating an oil spill. Black represents oil and white water; (b)Simulated SAR image}
  \label{fig:EM_forArticle_1}
\end{figure}

\begin{figure}
  \centering
  \includegraphics[width=70mm,angle=0]{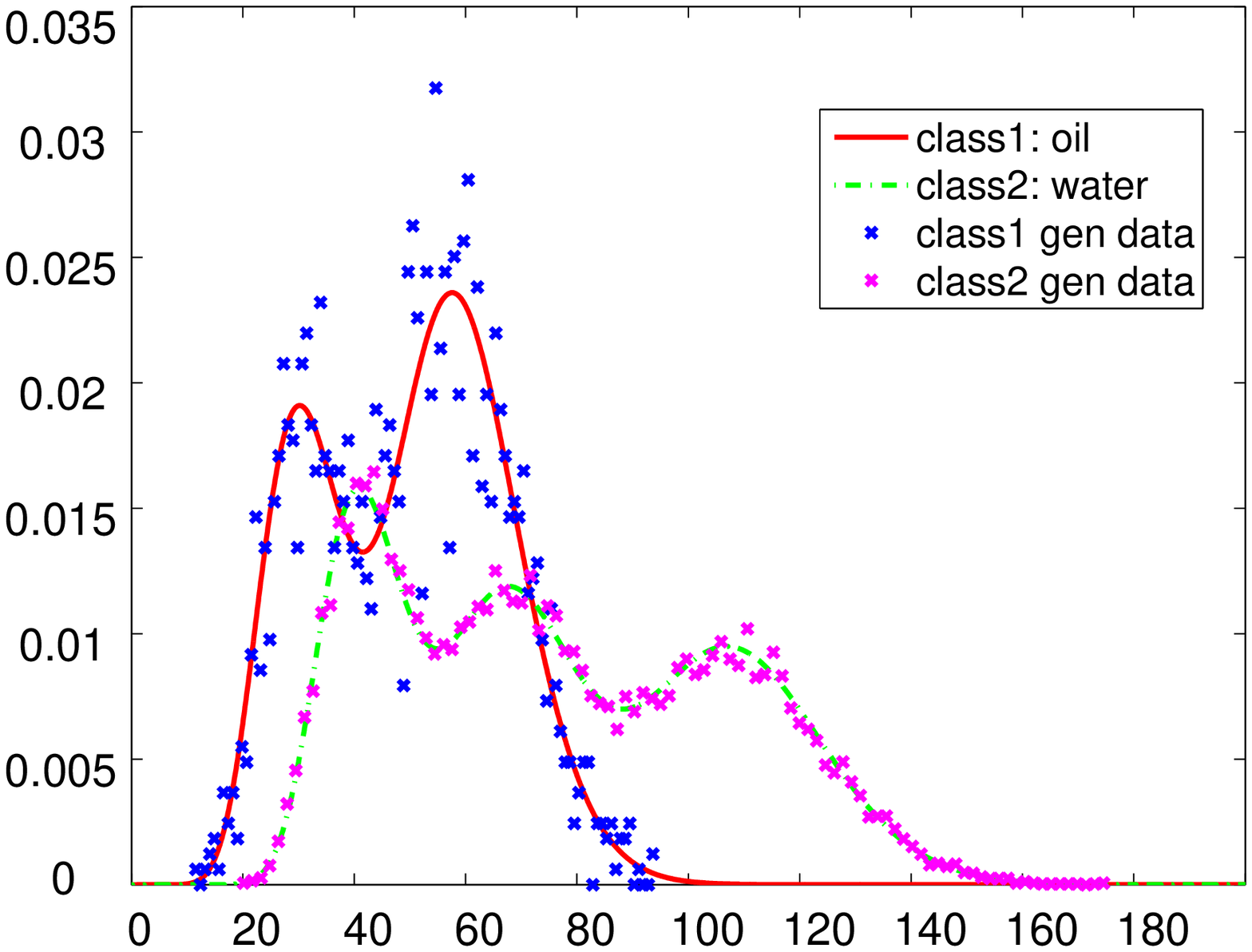}
  \caption{Probability functions used to generate the simulated image with superimposed histogram of generated data set. A three modes function for water and two modes function for oil was used.}
  \label{fig:EM_forArticle_2}
\end{figure}

\begin{figure}
  \centering
  \includegraphics[width=70mm,angle=0]{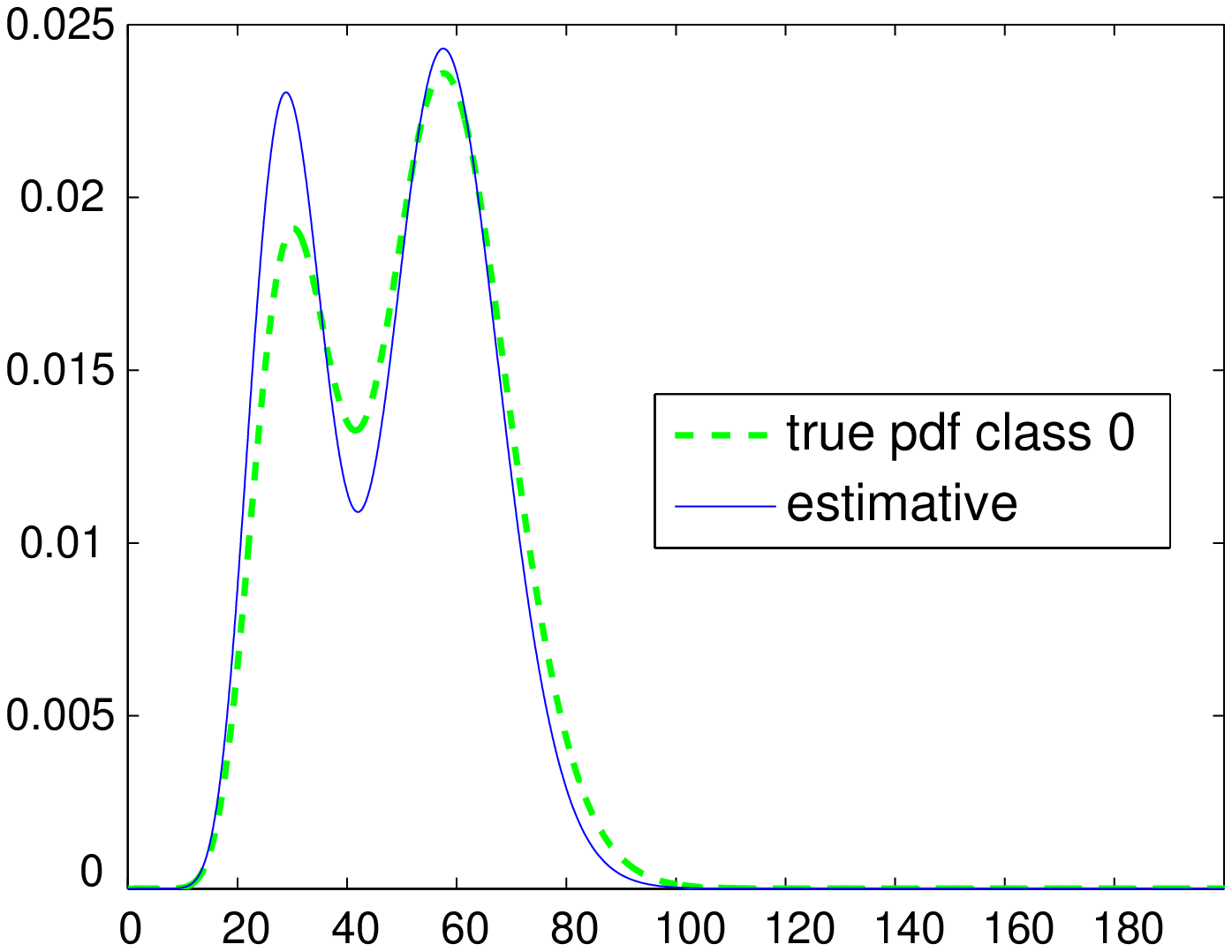}
  \caption{True and estimated class densities for oil.}
  \label{fig:EM_forArticle_3}
\end{figure}

\begin{figure}
  \centering
  \includegraphics[width=70mm,angle=0]{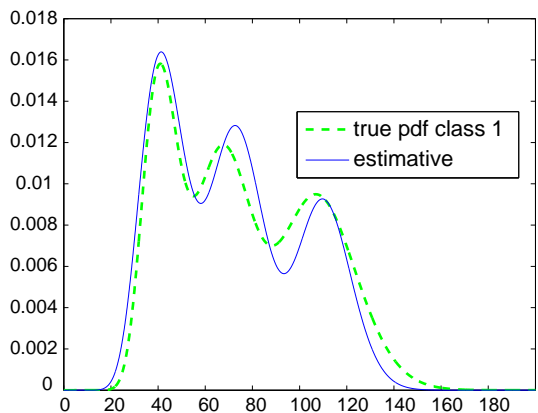}
  \caption{True and estimated class densities for water.}
  \label{fig:EM_forArticle_4}
\end{figure}

In this subsection we illustrate the behavior of the EM algorithm designed to infer Gamma mixtures. For details on the theoretical issues, we refer to the Appendix. In Fig. \ref{fig:EM_forArticle_1}, we can see the ground-truth used for simulating the image and the generated image, according to the densities shown in Fig. \ref{fig:EM_forArticle_2}. The two classes, oil and water have been modeled respectively by a mixture of two Gamma functions and a mixture of three Gamma functions. For this particular case, we have selected a ROI containing 178 pixels for representing the water and a ROI containing 142 pixels for representing the oil. After 20 iterations of the EM algorithm, we obtained the approximations for the probability distributions depicted in Fig. \ref{fig:EM_forArticle_3} and Fig. \ref{fig:EM_forArticle_4}. The obtained results are quite reasonable for the small sample sizes used.

\subsection{Real Images}\label{sec:6.2}
We have tested the proposed methodology with three real SAR images for the special purpose of oil spill detection. For this type of application, an unsupervised algorithm is more indicated, and so we decided to apply Algorithm 3. Because Loopy Estimation has proven to be an effective method in the simulations, we have choosen Algorithm 3 with Loopy Estimation. Nevertheless, in order to have a comparision unsupervised approach versus supervised approach, we also applied Algorithm 2 to two of the images.
In oil spill detection, the number of classes is typically set to $c=2$, although more classes can be considered if we are interested in distinguishing other phenomena occuring at the same time in the region of interest. The application of the algorithms is straightforward: the key idea, like in most state-of-the-art oil spill detection methods (see for example \cite{art:Solberg:OSDIR:07}), is to partition the image in tiles and run the algorithm separately for each part. This step is preceeded by the application of a landmask to the image, what can be done using external coastline information or by adopting some coastline self-extraction procedure. After the segmentation of each tile, a procedure for grouping patches detected on the tile borders should be carried on. Another possibility to increase segmentation coherency in the borders is to define overlapping tiles or to force continuity to some degree on the estimated class and/or smoothness parameters from one tile to the next. Nevertheless, at the moment we are not doing this and these are considered future possible improvements to our methodology.
We segmented oil spills contained in three different scenes, described in the  following subsections.

\subsubsection{Segmentation of  an ERS-1 sub-scene}\label{sec:6.2.1}
We have segmented part of an ERS-1 image from the Sicily Channel, Italy, that has been acquired on the 30$^{\textrm{th}}$ January of 1992. This image is referred in the ESA web pages regarding oil slicks \url{http://earth.esa.int/ew/oil_slicks/} and contains three oil slicks, along with information regarding wind direction and intensity and existence of ships, ship wakes, natural oil films and currents.

\begin{figure}
  \centering
  \includegraphics[width=80mm,angle=0]{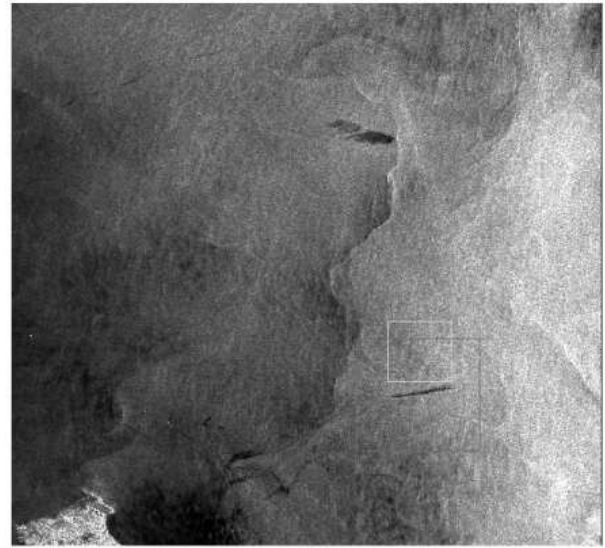}
  \caption{ERS-1 image from the Sicily Channel, Italy, acquired on 30th January of 1992. The smaller and larger squares are sub-scenes that have been used respectively to estimate the wind direction and to apply our algorithm.}
  \label{fig:RealSARImage}
\end{figure}

\begin{figure}
  \centering
  \includegraphics[width=85mm,angle=0]{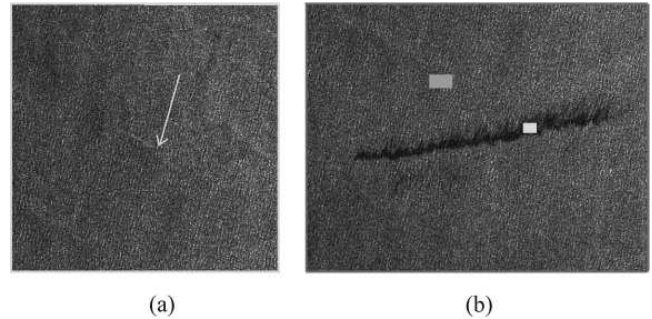}
  \caption{Closer look to the sub-scenes of the ERS-1 image. Part (a) is used for wind direction estimation; the estimated direction has been overlapped as a green arrow. Part (b) contains an oil spill. The coloured regions correspond to the ROI's selected for class parameter estimation in the supervised algorithm}
  \label{fig:ZOOMSofRealSarImage}
\end{figure}

\begin{figure}
  \centering
  \includegraphics[width=70mm,angle=0]{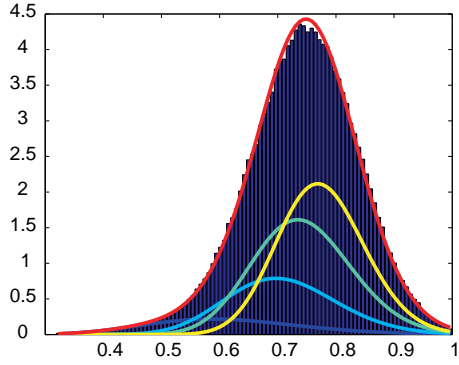}
  \caption{Fitting of a mixture of Gammas to the data from the red square in the ERS-1 image.}
  \label{fig:GammaFitInitialImageERS1}
\end{figure}

\begin{figure}
  \centering
  \includegraphics[width=70mm,angle=0]{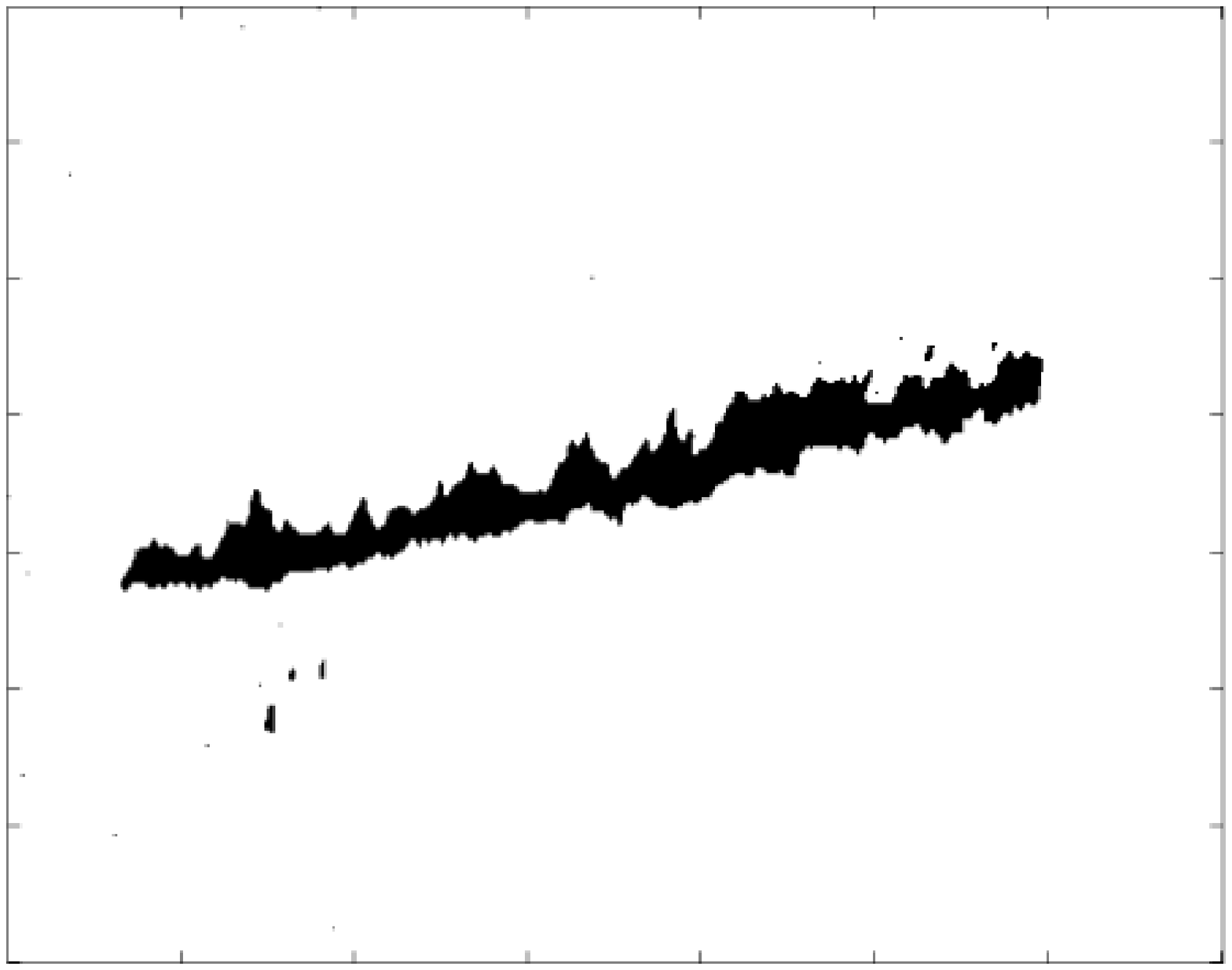}
  \caption{Segmentation of ERS-1 subscene containing oil with $\beta=1.4$, estimated using Algorithm-3 with the Loopy-$\beta$-Estimation method.}
  \label{fig:SegmentationERS1Alg3}
\end{figure}

\begin{figure}
  \centering
  \includegraphics[width=90mm,angle=0]{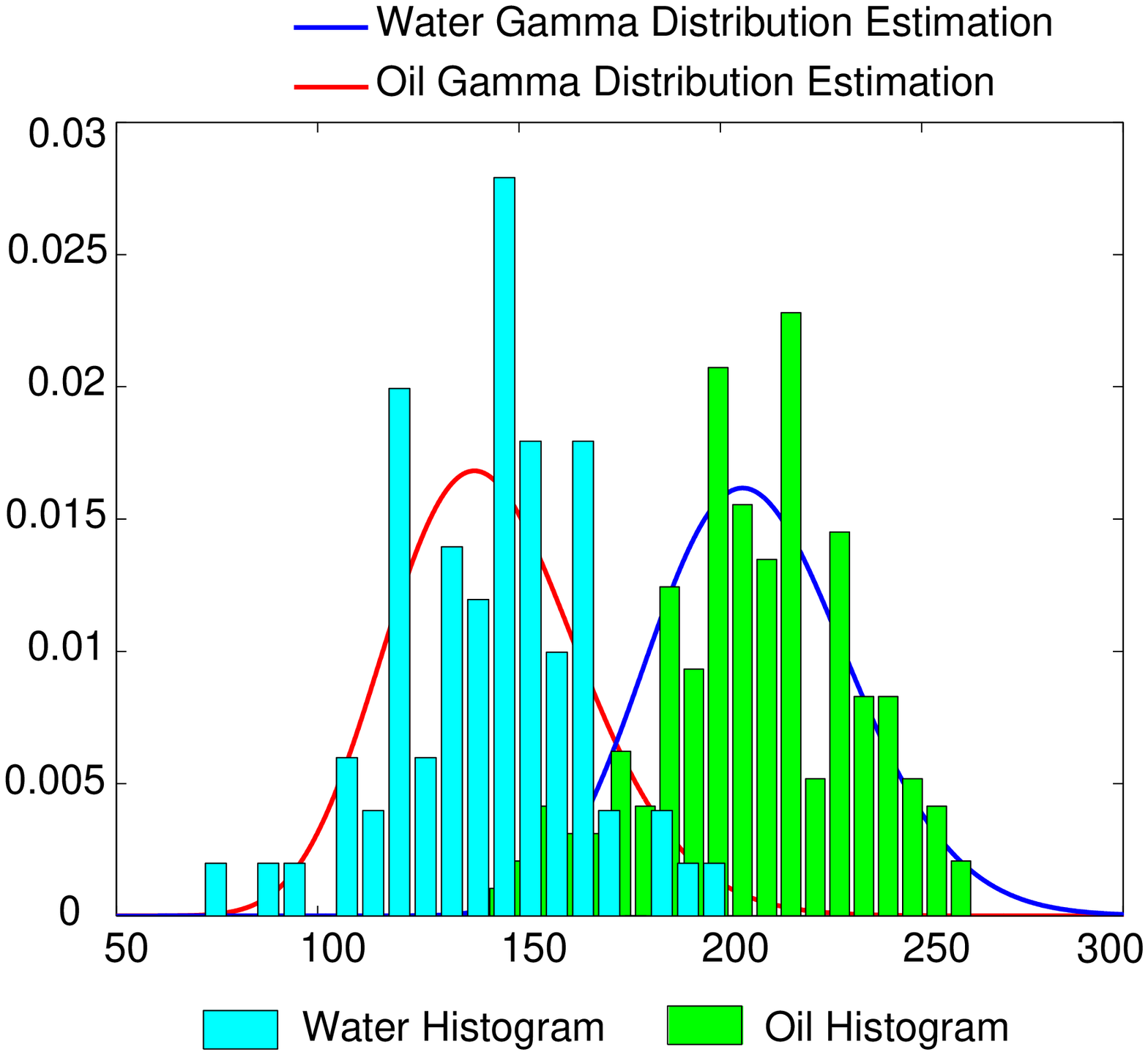}
  \caption{Class Parameters Estimation for Algorithm-2.}
  \label{fig:ClassParameters}
\end{figure}

\begin{figure}
  \centering
  \includegraphics[width=70mm,angle=0]{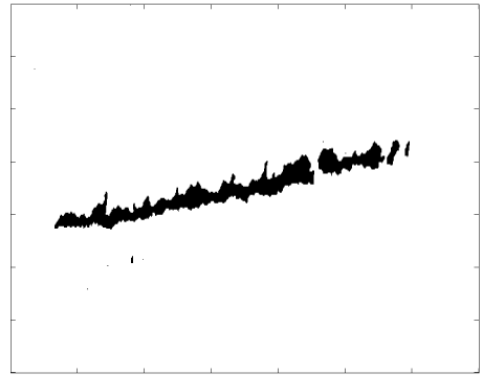}
  \caption{Segmentation of ERS-1 subscene containing oil with $\beta=1.44$, estimated using Algorithm-2.}
  \label{fig:SegmentationERS1Beta1Point5}
\end{figure}

Fig. \ref{fig:RealSARImage} provides a quicklook of the scene with two squares superimposed: the larger representing the part to be segmented and the smaller representing a part used for wind estiamtion. Fig. \ref{fig:ZOOMSofRealSarImage} provides zooms of the referred squares. By applying the Radon Transform to the smaller square, the estimated direction has been calculated and is depicted in the image. The direction was consistent with the measured value reported in the url site and is at the origin of the well-known "feathering" effect that can be observed in this linear spill.
We have computed the backscattering values of the image, by performing calibration using the ESA provided BEST software (\url{http://earth.esa.int/services/best/}) and then applied Algorithm-3 with Loopy-$\beta$-Estimation. The result of the Gamma mixture estimation, in the initialisation of the algorithm, is depicted in Fig. \ref{fig:GammaFitInitialImageERS1}. After only three iterations, both the class parameters and the smoothness parameter have converged. A $\beta$ value equal to 1.4 was estimated. The segmentation is displayed in Fig. \ref{fig:SegmentationERS1Alg3}. We also applied Algorithm-2: we selected two ROI's (160 pixels for water and 77 for oil) in the image (shown in Fig. \ref{fig:ZOOMSofRealSarImage}) and computed ML Gamma Estimators for the two classes (see Fig. \ref{fig:ClassParameters}). We then applied the algorithm and obtained an estimated $\beta$ value equal to 1.44. The segmentation result is given in Fig. \ref{fig:SegmentationERS1Beta1Point5} and is similar to the result obtained with Algorithm-3.

\begin{figure}
  \centering
  \includegraphics[width=60mm,angle=0]{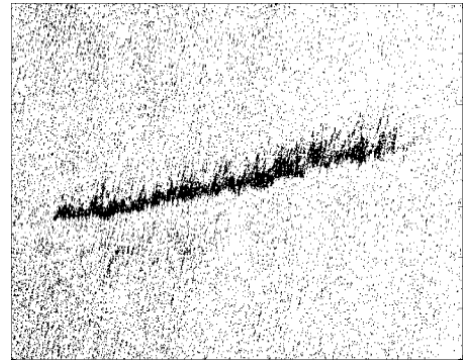}
  \caption{Segmentation of ERS-1 subscene containing oil, with $\beta=0$.}
  \label{fig:SegmentationERS1Beta0}
\end{figure}

For comparison, we also provide the segmentation obtained with no prior (corresponding to $\beta = 0$) in Fig. \ref{fig:SegmentationERS1Beta0}.
In practice, the estimated value has proved to deliver a good segmentation. When lower $\beta$ values were used, the result was noisier and for higher $\beta$ values, the details of the spill disappeared.

\subsubsection{Segmentation of an Envisat ASAR IM sub-scene}\label{sec:6.2.2}
\begin{figure}
  \centering
  \includegraphics[width=90mm,angle=0]{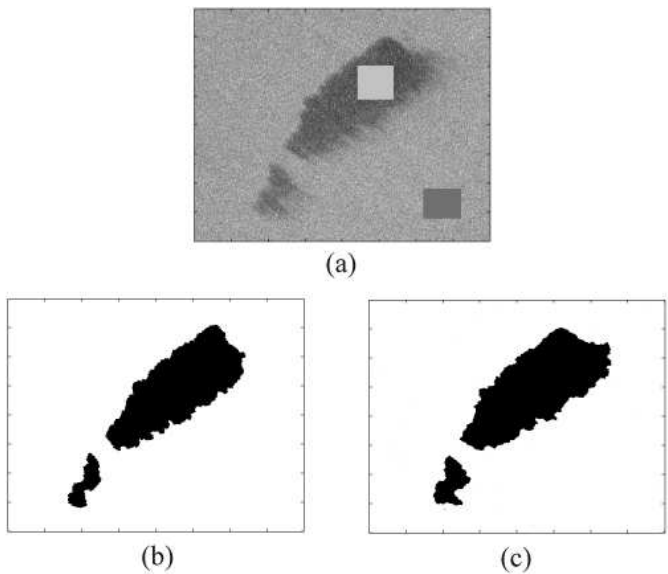}
  \caption{(a) ASAR image fragment with coloured regions for class parameter estimation in Algorithm-2; (b) Segmentation result applying Algorithm-2; (c) Segmentation result applying Algorithm-3 with Loopy-$\beta$-Estimation.}
  \label{fig:Alg3Real.eps}
\end{figure}

We have also applied Algorithm 3 to a fragment of an ASAR Image Mode image, acquired on 19 July 2004, in the ocean between Cyprus and Lebanon. The fragment contained
an occurred oil spill of circa 10 km's (centered on $\approx 33^\circ$N$, 33^\circ$E$39$') that was documented on the EC Oceanides project (a project in the framework of the Europe 's Global Monitoring for Environment and Security initiative) database. After six iterations, we achieved convergence of the $\beta$ and $\theta$ values. The segmentation result is given in Fig. \ref{fig:Alg3Real.eps}, corresponding to a estimated $\beta = 1.83$. The unsupervised segmentation result is compared with the one provided by Algorithm-2, where the user provided ROI's for water and for oil that were used to estimate the class parameters at the beginning of the process (see Fig. \ref{fig:Alg3Real.eps}). In this case a $\beta=1.75$ was obtained.
\\
\subsubsection{Segmentation of an Envisat ASAR WSM sub-scene}\label{sec:6.2.3}
\begin{figure}
  \centering \includegraphics[width=80mm, height=80mm, angle=0]{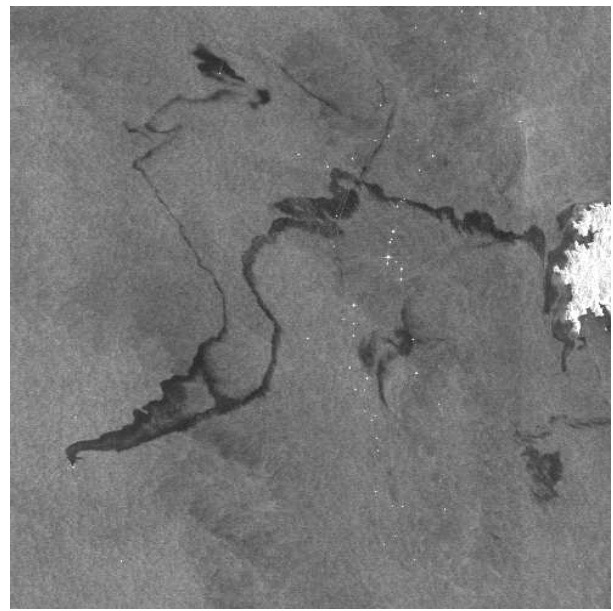}
  \caption{Display of the ASA WSM full resolution image of the Prestige accident occured in November 2002 in Galicia.}
  \label{fig:Alg3RealPrestige1.eps}
\end{figure}

\begin{figure}
  \centering
  \includegraphics[width=80mm,height= 80mm, angle=0]{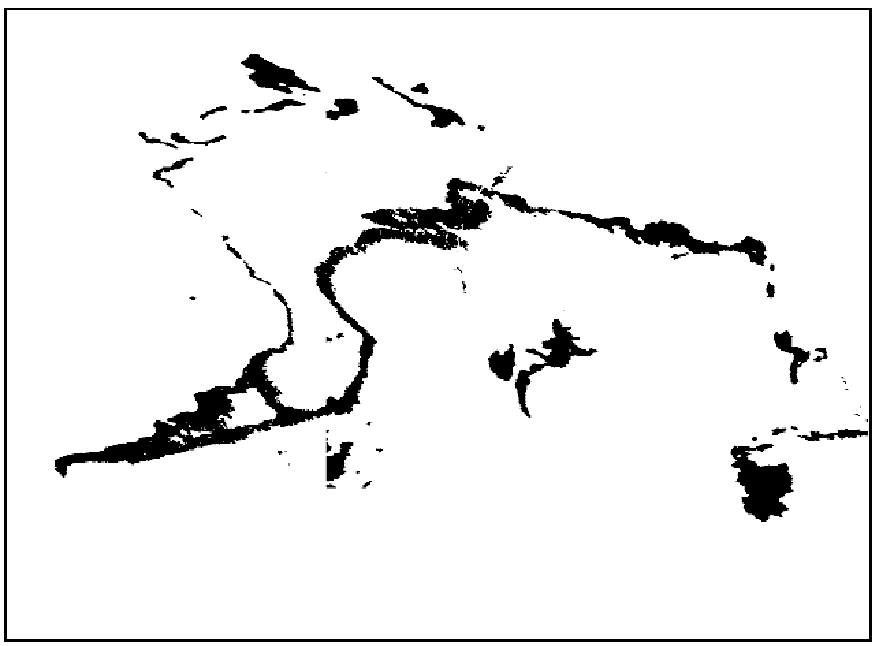}
 \caption{Display of the segmentation results of applying Algorithm 3 with Loopy-$\beta$-Estimation to the image displayed in Fig. \ref{fig:Alg3RealPrestige1.eps}.}
  \label{fig:Alg3RealPrestige2.eps}
\end{figure}

\begin{figure}
  \centering
  \includegraphics[width=70mm, angle=0]{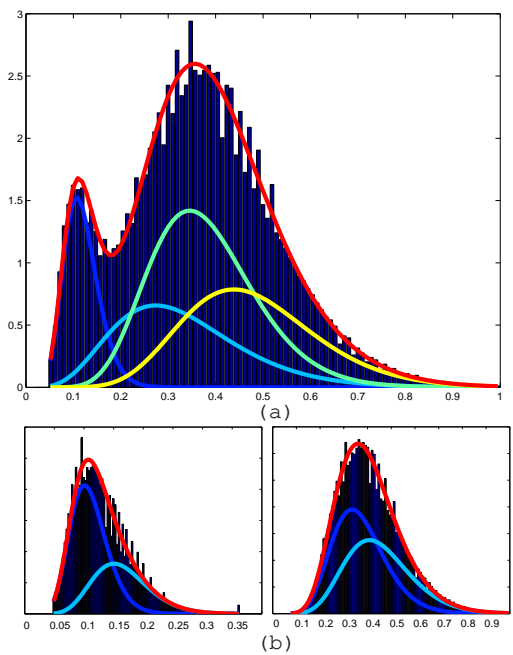}
  \caption{Results of applying the EM Gamma mixture estimation procedure to the tile depicted in Figure \ref{fig:tile}: (a) Histogramm of the data values (normalised) with superimposed estimated Gamma mixture; (b) Distributions corresponding to the oil class (left) and water class (right) after nine iterations of Algorithm 3: Histogramm and superimposed fitting.}
  \label{fig:Alg3RealPrestige4.eps}
\end{figure}

\begin{figure}
  \centering
  \includegraphics[width=70mm, height= 60mm, angle=0]{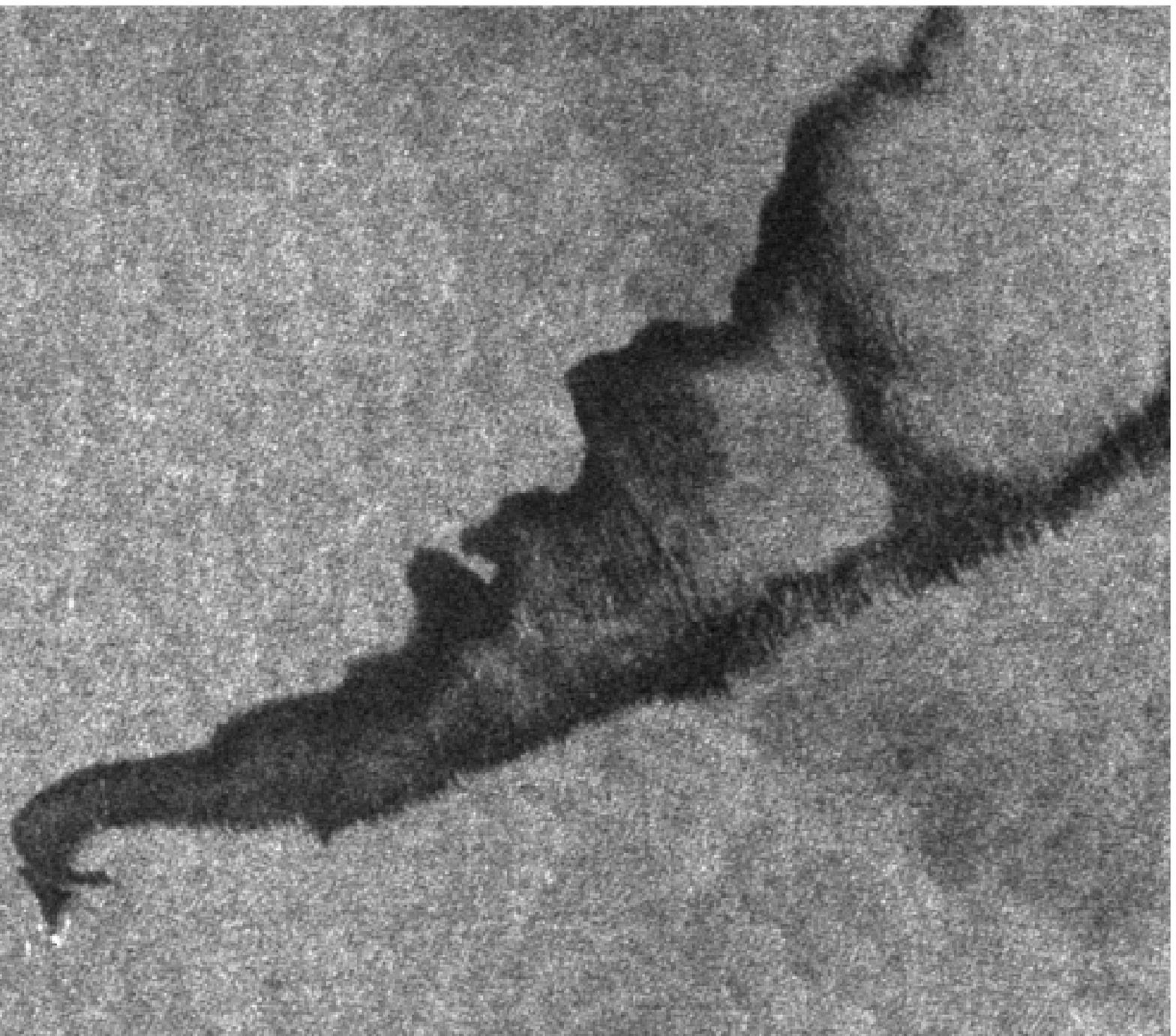}
  \caption{Display of one of the individual tiles in which the image in Figure \ref{fig:Alg3RealPrestige1.eps} has been divided for segmentation.}
  \label{fig:tile}
\end{figure}

In order to demonstrate the viability of applying Algorithm 3 to a whole ASAR\_WSM scene, we have run it over the very well known image of a confirmed oil spill, namely the Prestige case (see Fig. \ref{fig:Alg3RealPrestige1.eps}). This accident took place in November 2002, in Galicia (Spain), when a
tanker carrying more than 20 million gallons (around 67,000 tons) of oil split in half off the northwest coast of Spain on 19 November 2002, threatening one of the worst environmental disasters in history. For segmenting the image, we have first partitioned it in tiles of 600x600 pixels each and
then applied Algorithm-3 to each tile independently. No post-processing of the borders or ``clean-up'' operations (like for example morphological operations) have been carried out. Fig. \ref{fig:Alg3RealPrestige2.eps} shows the segmented image, made up of the concatenation of the segmentation of the individual tiles. As we can see, the result can be considered in general very good, with only some discrepancies located on the tiles' borders.
For initialising the algorithm, the EM Gamma mixture estimation procedure was applied with four modes ($K=4$). In fact this number is enough to provide a good fitting of the intensity data of the SAR image. On the other hand, our EM algorithm allows us to start with a higher mode number, and decreases this number automatically. We depict the results of this fitting on one of the tiles, shown in Figure \ref{fig:tile}, by showing the histogram of the tile with the superimposed estimated Gamma mixture (see Fig. \ref{fig:Alg3RealPrestige4.eps} (a)). As we have explained, the mode corresponding to the lower mean value is assigned to the oil class and the others to the water class, providing an initialisation to the class parameters. The estimations of these parameters are actualised along the algorithm and, after nine iterations, we obtain the distributions shown in Fig. \ref{fig:Alg3RealPrestige4.eps} (b).

\begin{figure}
  \centering
  \includegraphics[width=70mm,angle=0]{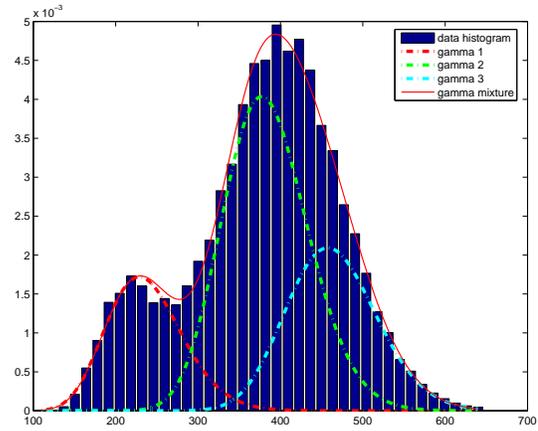}
  \caption{Histogramm of the data values (normalised) with superimposed estimated Gamma mixture when the number of classes is set to $c=3$}
  \label{fig:Alg3RealPrestige5.eps}
\end{figure}

\begin{figure}
  \centering
  \includegraphics[width=70mm, height= 60mm, angle=0]{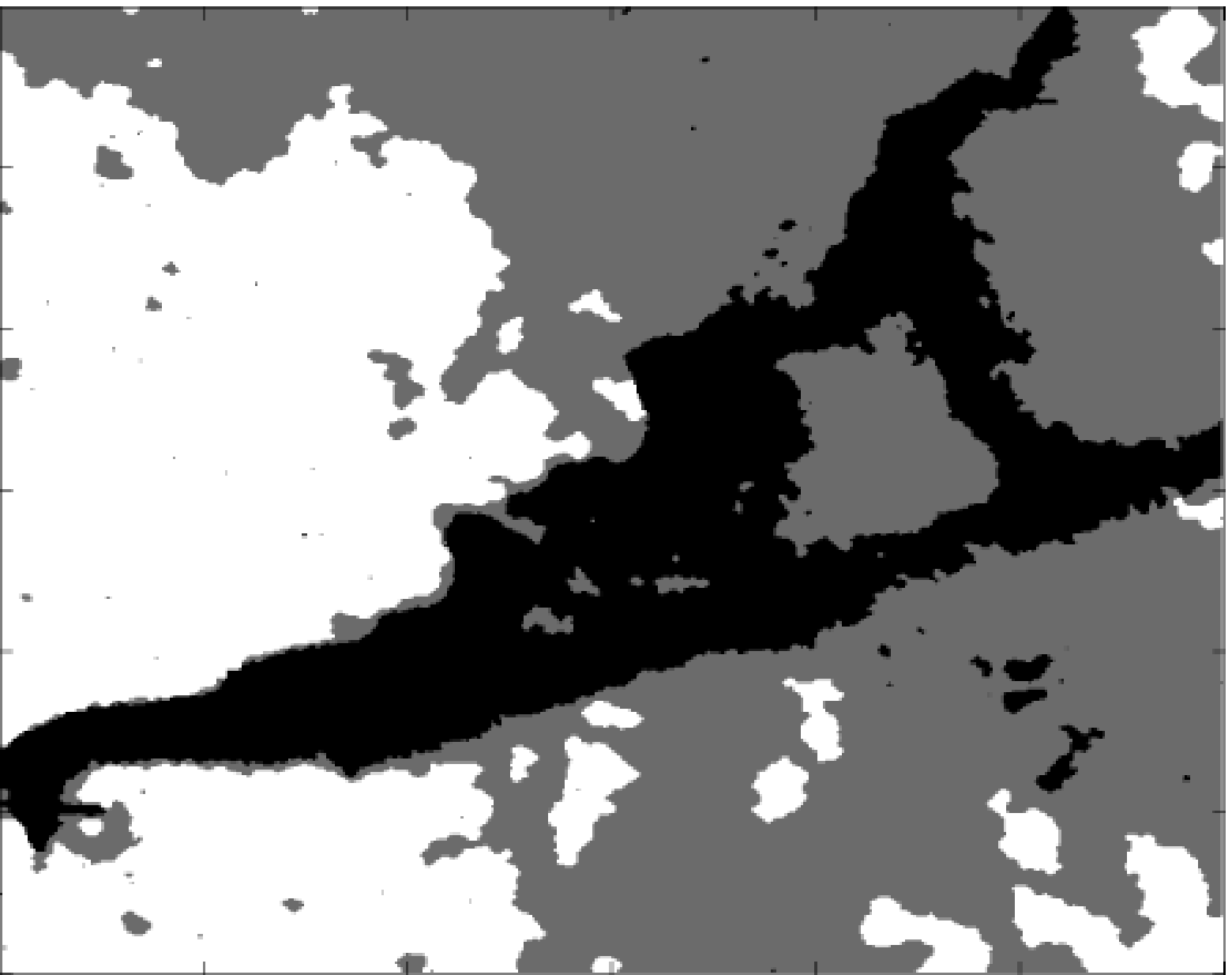}
  \caption{Display of the initial segmentation result of applying Algorithm 3 with
Loopy-$\beta$-Estimation and three classes to the tile depicted in Figure \ref{fig:tile}.}
  \label{fig:Alg3RealPrestige6.eps}
\end{figure}

\begin{figure}
  \centering
  \includegraphics[width=70mm, height= 60mm, angle=0]{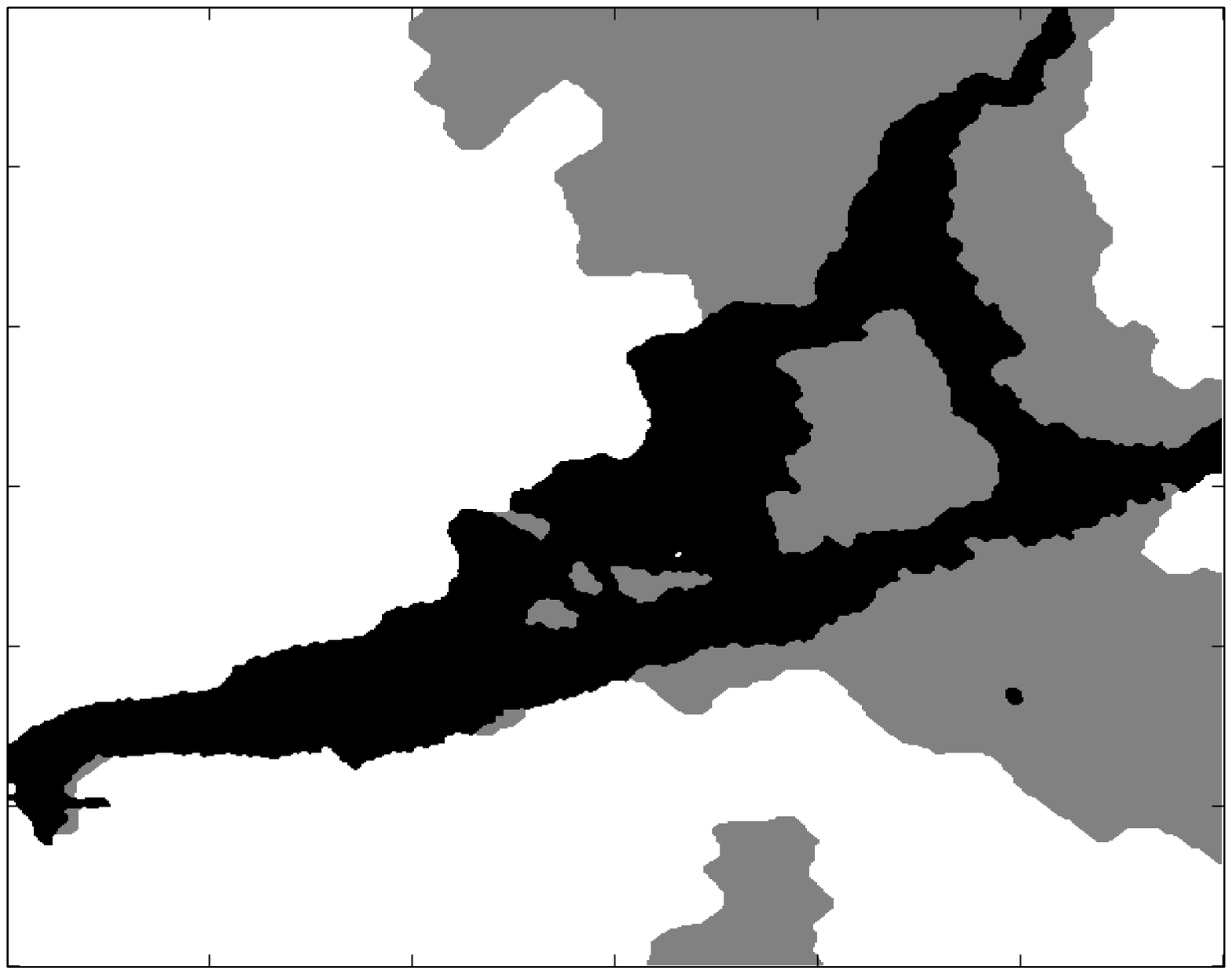}
  \caption{Display of the final segmentation result, after nine iterations, of applying Algorithm 3 with
Loopy-$\beta$-Estimation and three classes to the bottom left tile of Fig. 17.}
  \label{fig:Alg3RealPrestige7.eps}
\end{figure}

To fully demonstrate the possibilities of our algorithm, we again run it on the tile shown in Figure \ref{fig:tile}, but this time setting the number of classes to $c=3$. By doing so, we hope to be able to segment a third ambiguous zone, corresponding to intermediate radiometry levels and probably due to atmospheric conditions originating a front. In this case we choose to fitt a Gamma mixture of $c$ modes to the data for initialising the class parameters, as depicted in Fig. \ref{fig:Alg3RealPrestige5.eps}. With this initialisation, the first obtained segmentation using Algorithm 3 is displayed in Fig. \ref{fig:Alg3RealPrestige6.eps}, after only nine iterations, all parameters have already converged to the segmentation given in Fig. \ref{fig:Alg3RealPrestige7.eps}. When comparing this segmentation with the one provided by a state-of-the-art algorithm, namely by multiscale HMC model in \cite{art:Derrode:PR:07}, we consider to have obtained a very good result.

\subsubsection{Time considerations}\label{sec:6.3}
The proposed algorithms display, in practical cases, a computational complexity of $O(N)$, with N being the number of pixels in the image. In fact, the most time-consuming steps in the algorithms are the $\alpha$-Expansion and Graph-Cut segmentation routines. These use the max-flow code implementation referred in \cite{art:Boykov:K:PAMI:04} that has complexity $O(n^{2.5})$ in the worst-case, however, in most situations is $O(N)$.
Both Algorithm 1 and Algorithm 3 converge in few iterations and, as an example, for images of the dimension used in the first simulation (64x64) (see \ref{sec:6.1.1}), only a few seconds are needed to run the algorithms. For a 600x600 pixel tile, like those in the Prestige example in \ref{sec:6.2.3}, Algorithm 3 is taking $\approx$ 1 minute, but this time performance could be improved by a more efficient implementation of our procedure, namely by improving the implementation of the Loopy-$\beta$-Estimation and by migrating our matlab code fully to C-code.

% File Name         : OILSAR_article_sec7.tex
% Contents          : sec7 (two columns)
% Paper Title       : Oil Spill Segmentation
% Paper ID          :
% Person ID         :
% Date              : 8/7/05
% Author            : Jose M. Bioucas-Dias and S\'{o}nia Pelizzari

\section{Conclusions}

The results of applying the proposed methodology to simulated images with Gamma data models and to real SAR data are promising.
The developed EM Gamma mixture estimation algorithm, when incorporated into the proposed segmentation algorithms, has proved to be an efficient tool for data modeling in SAR intensity images.

With the supervised algorithms, high OA accuracies have been achieved even for simulated images with high levels of noise. In general the Bayesian approach has resulted in an OA increase in the segmentation process between $10$ and $15\%$, when compared to the segmentation using no prior information. Hereby, Algorithm 1 with LSF and Algorithm 1 with CD methods provided similar results, with a performance close to that obtained by setting the $\beta$ value manually. On the other hand, the LSF estimation procedure seems to become less reliable for noisier images. When compared to Algorithm 2, using Loopy-$\beta$-Estimation, we see that this last method allways provides equal or better results, outcoming Algorithm 1.
Another advantage of Algorithm 2 is that it is usually faster than Algorithm 1, being a one-shot process. Both algorithms, by introducing prior information into the segmentation process, increase the OA significantly. As a conclusion, Algorithm 2 should be preferred to Algorithm 1, when using a supervised method for oceanic sar images segmentation.

Algorithm 3, totally unsupervised, has been conceived as an improvement to Algorithm 1 and 2. When testing it on simulated data the obtained results were very good, although slightly worst than the supervised ones, as expected. In general, after a few iterations, the class and smoothness parameters converge to stable and meaningful values.
By applying Algorithm 3 to real images containing documented oil spills, the segmentation has been considered successfull. We could segment both linear and patch type oil spills. Furthermore, the applicability of the method to segment whole scenes, as well as to segment more than two classes, has also been demonstrated in a well known image from the Prestige accident.
As a conclusion, we believe the presented methods are suitable to be used for segmenting oceanic SAR images. In particular for oil spill detection, Algorithm 3 seems to be a suitable method. 

Based on observations that oil spills in the ocean are often dragged by the wind and align more or less perpendicular to its direction, an interesting future issue is the incorporation of wind information into our segmentation algorithms. By adopting anisotropic MRF in the prior, we intend to reflect this directional dependency of the clique potentials. In the practice, the smoothness parameter $\beta$ is no longer considered to be a constant value but is clique-dependent according to the wind direction and/or velocity. For estimating the wind profile from the SAR data, state-of-the-art algorithms are used. We have run promissing simulations, are currently testing this extension on real data, and expect to obtain interesting results on a near future.

\appendix

\section*{\bf Fitting a Mixture of Gamma Densities}

As stated in the main text of this work, when segmenting real  SAR images, the
adopted data model for each class is a finite Gamma mixture. For completely
defining the data model for each class (for lightness of the notation we have
dropped the class index),  we need to estimate $2K$ Gamma parameters, $\theta =
(\theta_1,\dots,\theta_K)$, with $\theta_s = (a_s,\lambda_s)$, and $K$ a priori
probabilities $\alpha=(\alpha_1,\dots,\alpha_K))$. We infer
$\phi=(\alpha,\theta)$ by computing its ML estimate from a training set. The ML
estimate is computed via an EM algorithm \cite{art:Dempster:MLE_EM:77}. Fitting
a Gamma mixture is addressed in \cite{conf:Zaart:VI:99}.  However, the authors
consider  $N$-look SAR images meaning that the underlying random variables are
the average of $N$ independent and identically distributed exponential random
variables, thus,  having a Gamma density but with just one parameter free; if the
mean is $\mu$, then the variance is given by $\mu^{2}/N$. We estimate both the
mean and the variance for each Gamma distribution in the mixture, rendering the
algorithm more adaptable to real measurements.

The key point in the EM technique is the introduction of the so called missing
data $z$, such that $p\left(y|\phi\right)=\int p\left(y,z|\phi\right)dz$ and
$p\left(y,z|\phi\right)$ is easier to manipulate than $p\left(y|\phi\right)$.
In the particular case of a mixture of densities, we will use as missing data a
random variable, $z_{i}$, per site, with distribution $p(z_{i}=s)=\alpha_{s}$.
It is interpretable as  the probability of the $s-th$ Gamma mode is selected
at pixel $i$. The EM algorithm alternates between two steps: the E-step
computes the conditional expectation of the logarithm of the complete a
posteriori probability function, with respect to the missing variables, based
on the actual parameter value. The M-step updates the values of the parameters,
by maximizing the expression obtained in the E-step with respect to each
parameter on turn, \emph{i.e.},

\begin{eqnarray}
\mbox{E\--step}& :& Q\left(\phi;\phi^{t}\right) =
E\left\{\log{p\left(y,z|\phi\right)|y,\phi^{t}}\right\}
\label{eq:EM_GammaMix3}\\
\mbox{M\--step}& :& \phi^{t+1} = \arg\max_{\phi}Q\left(\phi;\phi^{t}\right)
\label{eq:EM_GammaMix4}
\end{eqnarray}
Denoting
\begin{equation}
w_{si}^{t}=P\left(z_{i}=s|y_{i},\phi^{t}\right)
\label{eq:EM_GammaMix5}\\
\end{equation}
and taking into account that $\sum\alpha_{s}$ = 1,  then $\phi^{t+1}$ can be
found among the stationary points of the Lagrangean
\begin{equation}
{\cal L}(\phi)=
\sum_{i=1}^{N}\sum_{s=1}^{K}\left(L_{si}(\theta_s,\alpha_s)\,w_{si}^{t}\right)+
\lambda\left(\sum_{s=1}^{K}\alpha_{s}-1\right),
\label{eq:EM_GammaMix6}\\
\end{equation}
with
\begin{eqnarray}
L_{si}(\theta_s,\alpha_s)& =& \log{\left(\lambda_{s}^{a_{s}}\right)}-
\log{\left[\Gamma\left(a_{s}\right)\right]}\nonumber\\
     && +\log{\left(y_{i}^{a_{s}-1}\right)} -\lambda_{s}y_{i}  +\log{\left(\alpha_{s}\right)},
\label{eq:EM_GammaMix7}
\end{eqnarray}
where $\lambda$ denotes a Lagrange multiplier. The expression  for $w_{si}^{t}$
(see \cite{book:McLachlanand:FMM}) is given by
\begin{equation}
w_{si}^{t}=\frac{\alpha_{s}^tp(y_{i}|\theta_{s}^{t})}{\sum_{r=1}^{K}\alpha_{r}^tp(y_{i}|~\theta_{r}^{t})}.
\label{eq:EM_GammaMix8}\\
\end{equation}
In the M-step, after differentiating ${\cal L}$ in order to the  unknown
parameters and setting the derivatives to zero, we obtain a closed solution for
the updating of the a priori probabilities $\alpha_{i}$'s, but numerical
iteration is needed for determining parameters $a_{i}$'s and $\lambda_{i}$'s of
the Gamma densities. Expression (\ref{eq:EM_GammaMix9}) gives the update
expression for $\alpha_{i}$'s.
\begin{equation}
\alpha_{s}^{t+1}=\frac{1}{N}\sum_{i=1}^{N}w_{si}^{t}.
\label{eq:EM_GammaMix9}\\
\end{equation}
Equation (\ref{eq:EM_GammaMix10}) and (\ref{eq:EM_GammaMix11})  give the update
expressions for the parameters $\lambda_{i}$'s and $a_{i}$'s.
\begin{equation}
\lambda_{s}^{t+1}=\frac{a_{s}^t\sum_{i=1}^{N}w_{si}^t}{\sum_{i=1}^{N}y_{i}w_{si}^t},
\label{eq:EM_GammaMix10}\\
\end{equation}
\begin{equation}
a_{s}^{t+1}=\Psi^{-1}\left[\frac{\log{\left(\lambda_{s}^t\right)}\sum_{i=1}^{N}w_{si}^t
                   +\sum_{i=1}^{N}\log{(y_{i})}w_{si}^t}{\sum_{i=1}^{N}w_{si}^t}\right],
\label{eq:EM_GammaMix11}\\
\end{equation}
where
\begin{equation}
\Psi\left(a_{s}\right)=\frac{\Gamma'\left(a_{s}\right)}{\Gamma\left(a_{s}\right)},
\label{eq:EM_GammaMix12}\\
\end{equation}
is the psi function. We refer to the Appendix B  of \cite{thesis:Nascimento:06}
for a very fast Newton procedure to compute the inverse of the psi$(\cdot)$ function.
Expressions (\ref{eq:EM_GammaMix10}) and (\ref{eq:EM_GammaMix11}) are
iteratively recomputed until convergence is obtained, starting from initial
values computed from the observed data y. The initial parameter values are
calculated in such a way, that the initial probability function is a sum of
equidistant Gammas that span the most representative data range. The EM scheme
converges in a few tens of iterations.

\section*{Acknowledments}

We thank V. Kolmogorov for the max-flow code made available  at
\url{http://www.cs.cornell.edu/People/vnk/software.html}, used in the Graph-Cut
and $\alpha$-Expansion segmentation techniques and S. Kumar for the Loopy
Belief Propagation code used for computing the two-node beliefs in the
Loopy-$\beta$-Estimation EM algorithm described in Section \ref{sec:3.3}.

\bibliographystyle{IEEE}
%\bibliography{sonia}

\begin{thebibliography}{10}

\bibitem{book:NOAA:seamanual}
C.~R. Jackson and J.~R. Apel,
\newblock {\em Synthetic Aperture Radar Marine User's Manual},
\newblock Commerce Dept., NOAA, 2005.

\bibitem{art:Wu:IJRS:03}
S.~Y. Wu and A.~K. Liu,
\newblock ``Towards an automated ocean feature detection, extraction and
  classification scheme for {SAR} imagery,''
\newblock {\em International Journal Remote Sensing}, vol. 24, no. 5, pp.
  935--951, 2003.

\bibitem{art:Brekke:OSD:07}
C.~Brekke and A.~Solberg,
\newblock ``Oil spill detection by satellite remote sensing,''
\newblock {\em Remote Sensing of Environment}, vol. 95, pp. 1--13, 2005.

\bibitem{art:Mercier:TGRS:06}
G.~Mercier and F.~Girard-Ardhuin,
\newblock ``Partially supervised oil-slick detection by {SAR} imagery using
  kernel expansion,''
\newblock {\em IEEE Transactions on Geoscience and Remote Sensing}, vol. 44,
  no. 10, pp. 2839--2846, February 2006.

\bibitem{conf:Pelizzari:SeaSAR:06}
S.~Pelizzari and Jos\'{e} M.~B. Dias,
\newblock ``Bayesian adaptive oil spill segmentation of {SAR} images via graph
  cuts,''
\newblock in {\em Proceedings of the SeaSAR}, 2006.

\bibitem{conf:Pelizzari:IbPRIA:07}
S.~Pelizzari and Jos\'{e} M.~B. Dias,
\newblock ``Bayesian oil spill segmentation of {SAR} images via graph cuts,''
\newblock in {\em Proceedings of the IbPRIA}, 2007, vol.~2, pp. 637--644.

\bibitem{conf:Pelizzari:IGARSS:07}
S.~Pelizzari and Jos\'{e} M.~B. Dias,
\newblock ``Oil spill segmentation of {SAR} images via graph cuts,''
\newblock in {\em Proceedings of the IGARSS}, 2007, pp. 1318 -- 1321.

\bibitem{conf:Bioucas:ICIP:99}
T.~Silva J.~B.~Dias and J.~Leit\ {ao},
\newblock ``Adaptive restoration of speckled {SAR} images using a compound
  random {M}arkov field,''
\newblock in {\em Proceedings of ICIP}, 1998, vol.~2, pp. 79--83.

\bibitem{art:Delignon:1997}
R.~Garello Y.~Delignon and A.~Hillion,
\newblock ``Statistical modelling of ocean {SAR} images,''
\newblock {\em IEEE Proc.-Radar, Sonar Navigation}, vol. 144, no. 6, pp.
  348--354, December 1997.

\bibitem{book:Li:ComputerVision}
S.Z.Li,
\newblock {\em Markov Random Field Modeling in Computer Vision},
\newblock Springer-Verlag, Tokyo, 1995.

\bibitem{conf:Yedidia:IJCAI:01}
J.~Yedidia, W.~Freeman, and Y.~Weiss,
\newblock ``Understanding belief propagation and and its generalizations,''
\newblock in {\em Proceedings of International Joint Conference on Artificial
  Intelligence}, 2001.

\bibitem{art:Kolmogorov:PAMI:04}
V.~Kolmogorov and R.~Zabih,
\newblock ``What energy functions can be minimized via graph cuts?,''
\newblock {\em IEEE Transactions on Pattern Analysis and Machine Intelligence},
  vol. 26, no. 2, pp. 147--159, February 2004.

\bibitem{art:Boykov:V:Z:PAMI:01}
Y.~Boykov, O.~Veksler, and R.~Zabih,
\newblock ``Fast approximate energy minimization via graph cuts,''
\newblock {\em IEEE Transactions on Pattern Analysis and Machine Intelligence},
  vol. 23, no. 11, pp. 1222--1239, 2001.

\bibitem{art:Derrode:PR:07}
S.~Derrode and G.~Mercier,
\newblock ``Unsupervised multiscale oil slick segmentation from {SAR} images
  using a vector hmc model,''
\newblock {\em Pattern Recognition}, vol. 40, pp. 1135--1147, 2007.

\bibitem{art:Maurizio:TGRS:07}
A.~Gambardella M.~Migliaccio and M.~Tranfaglia,
\newblock ``{SAR} polarimetry to observe oil spills,''
\newblock {\em IEEE Transactions on Geoscience and Remote Sensing}, vol. 45,
  no. 2, pp. 506--511, February 2007.

\bibitem{art:Solberg:OSDIR:07}
C.~Brekke A.~Solberg and P.Hus{\o}y,
\newblock ``Oil spill detection in {R}adarsat and envisat {SAR} images,''
\newblock {\em IEEE Transactions on Geoscience and Remote Sensing}, vol. 45,
  no. 3, 2007.

\bibitem{conf:Kanaa:IGARSS:03}
G.~Mercierb V.P. Onanac J. Mvogo Ngonoc P.L. Frisond J.P.~Rudantd T.~Kanaa,
  E.~Tonyea and R.~Garellob,
\newblock ``Detection of oil slick signatures in {SAR} images by fusion of
  hysteresis thresholding responses,''
\newblock in {\em Proceedings of IGARSS}, 2003, vol.~4, pp. 2750--2752.

\bibitem{art:Galland:SAR_OSS:04}
P.~R\'{e}fr\'{e}gier F.~Galland and O.~Germain,
\newblock ``Synthetic aperture radar oil spill segmentation by stochastic
  complexity minimization,''
\newblock {\em IEEE Geoscience and Remote Sensing Letters}, vol. 1, no. 4,
  2004.

\bibitem{art:Geman:IEEETPA:84}
S.~Geman and D.~Geman,
\newblock ``Stochastic relaxation, {G}ibbs distributions and the {B}ayesian
  restoration of images,''
\newblock {\em IEEE Trans. Patt. Analysis and Mach. Intel.}, vol. PAMI-6, no.
  6, pp. 721--741, 1984.

\bibitem{art:Berthod:IVC:96}
S.~YuE M.~Berthod, Z.~Kato and J.~Zerubia,
\newblock ``Bayesian image classification using {M}arkov random fields,''
\newblock {\em Image and Vision Computing}, vol. 14, pp. 285--295, 1996.

\bibitem{art:Kelly:TASSP:88}
H.~Derin P.~Kelly and K.D.Hartt,
\newblock ``Adaptive segmentation of speckled images using a hierarchical
  random field model,''
\newblock {\em IEEE Transactions on Acoustics, Speech, and Signal Processing},
  vol. 36, no. 10, pp. 1628--1641, 1988.

\bibitem{art:Derin:TGRS:90}
G.~V\'{e}zina H.~Derin, P.~Kelly and S.G.Labitt,
\newblock ``Modeling and segmentation of speckled images using complex data,''
\newblock {\em IEEE Transactions on Geoscience and Remote Sensing}, vol. 28,
  no. 1, pp. 76--87, 1990.

\bibitem{art:Dellepiane:TGRS:97}
P.C.Smits and S.G.Dellepiane,
\newblock ``Synthetic aperture radar image segmentation by a detail preserving
  {M}arkov random field approach,''
\newblock {\em IEEE Transactions on Geoscience and Remote Sensing}, vol. 35,
  no. 4, pp. 844--857, 1997.

\bibitem{book:Jakowatz96}
C.~Jakowatz, D.~Wahl, P.~Eichel, D.~Ghiglia, and P.~Thompson,
\newblock {\em Spotlight-Mode Synthetic Aperture Radar: A Signal Processing
  Approach},
\newblock Kluwer, Norwell, MA, 1996.

\bibitem{art:Jonhnson:Kotz:94}
N.~Jonhnson and S.~Kotz,
\newblock {\em Distribution in Statistics: Continuous Univarite Distributions},
  vol.~1,
\newblock Wiley, New York, 1996.

\bibitem{art:Dempster:MLE_EM:77}
N.~Laird A.~Dempster and D.~Rubin,
\newblock ``Maximum likelihood estimation from incomplete data via the {EM}
  algorithm,''
\newblock {\em Journal of the Royal Statistical Society B}, vol. 39, pp. 1--38,
  1977.

\bibitem{conf:Kumar:ISCV:95}
S.~Kumar, J.~August, and M.~Hebert,
\newblock ``Discontinuity preserving surface reconstruction through global
  optimization,''
\newblock in {\em EMMCVPR, LNCS}, 2005, vol. 3757, pp. 153--168.

\bibitem{art:Ibagnez:S:03}
M.~Ib{\'a}{\~n}ez and A.~Sim{\'o},
\newblock ``Parametric estimation in {M}arkov random fiels image modeling with
  imperfect observations. a comparative study,''
\newblock vol. 24, pp. 2377--2389, 2003.

\bibitem{art:Younes:89}
L.~Younes,
\newblock ``Parametric inference for imperfectly observed {G}ibbsian fields and
  some comments on {Chalmond's EM Gibbsian algorithm},''
\newblock {\em Probability Theory and Related Fields}, vol. 82, pp. 625--645,
  1989.

\bibitem{conf:Younes:92}
L.~Younes,
\newblock ``Parametric inference for imperfectly observed {G}ibbsian fields and
  some comments on {Chalmond's EM Gibbsian algorithm},''
\newblock in {\em Proc. Stochastic Models, Statistical Methods and Algorithms
  in Image Analysis}, P.~Barone and A.~Frigessi, Eds., Berlin, Germany, 1991,
  vol.~47 of {\em Lecture Notes in Statistics}, pp. 240--258, Springer.

\bibitem{art:Lakshamanan:D:89}
S.~Lakshamanan and H.~Derin,
\newblock ``Simultaneous parameter estimation and segmentation of gibbd random
  fields using simulated annealing,''
\newblock {\em IEEE Trans. Patt. Analysis and Mach. Intel.}, vol. 11, pp.
  793--813, 1989.

\bibitem{art:Boykov:K:PAMI:04}
Y.~Boykov and V.~Kolmogorov,
\newblock ``An experimental comparison of min-cut/max-flow algorithms for
  energy minimization in vision,''
\newblock {\em IEEE Transactions on Pattern Analysis and Machine Intelligence},
  vol. 26, no. 9, pp. 1124--1137, 2004.

\bibitem{conf:Zaart:VI:99}
S.~Wang Q.~Jiang A.~E.~Zaart, D.~Ziou and G.~B. B\'{e}ni\'{e},
\newblock ``{SAR} images segmentation using mixture of {G}amma distributions,''
\newblock in {\em Proceedings of Vision Interface}, 1999.

\bibitem{book:McLachlanand:FMM}
G.~McLachlanand and D.~Peel,
\newblock {\em Finite Mixture Models},
\newblock John Wiley and Sons, New York, 2000.

\bibitem{thesis:Nascimento:06}
Jos\'{e}~M. Nascimento,
\newblock {\em Unsupervised Hyperspectral Unmixing},
\newblock Ph.D. thesis, Instituto Superior Tecnico, 2006.

\end{thebibliography}

\end{document}